\documentclass[10pt,letter]{article}

\pdfoutput=1

\usepackage[english]{babel}
\usepackage{amsmath,amssymb}
\usepackage{dsdshorthand}
\usepackage{xcolor}
\usepackage{float}
\usepackage{booktabs}
\usepackage[utf8]{inputenc}

\definecolor{darkblue}{rgb}{0.1,0.1,.7}
\usepackage[colorlinks, linkcolor=darkblue, citecolor=darkblue, urlcolor=darkblue, linktocpage]{hyperref} 
\usepackage[square, comma, compress,numbers]{natbib}
   \usepackage[]{graphicx}
    \usepackage{geometry}
   \usepackage[margin=10pt,font=small,labelfont=bf]{caption}

\usepackage{titlesec}
\titleformat*{\section}{\large\bfseries}
\titleformat*{\subsection}{\normalsize\bfseries}
\titleformat*{\subsubsection}{\normalsize\it}
\titleformat*{\paragraph}{\normalsize\bfseries}
\titleformat*{\subparagraph}{\normalsize\bfseries}

\newcommand{\eps}{}

\def\eps{\epsilon}
\newcommand{\beq}{\begin{equation}} 
\newcommand{\eeq}{\end{equation}}
 
\def\nn{\nonumber}

\def\half{{\textstyle\frac 12}}

\def\ge{\geqslant}
\def\le{\leqslant}
\def\geq{\geqslant}
\def\leq{\leqslant}
\def\<{\langle}
\def\>{\rangle}

\numberwithin{equation}{section}
\usepackage[tocgraduated]{tocstyle}

\begin{document}

\vspace*{-.6in} \thispagestyle{empty}
\begin{flushright}
\end{flushright}
\vspace{1cm} 
{\Large
	\begin{center}
		{\bf Distributions in CFT I. Cross-Ratio Space}
\end{center}
}
\vspace{1cm}
\begin{center}
	{\bf Petr Kravchuk$^a$, Jiaxin Qiao$^{b,c}$,  Slava Rychkov$^{c,b}$}\\[2cm] 
	{
		$^{a}$ Institute for Advanced Study, Princeton, New Jersey 08540, U.S.A. \\
		$^b$  
		Laboratoire de Physique de l'Ecole normale sup\'erieure, ENS,\\ 
		{\small Universit\'e PSL, CNRS, Sorbonne Universit\'e, Universit\'e de Paris,} F-75005 Paris, France
		\\
		$^c$  Institut des Hautes \'Etudes Scientifiques, Bures-sur-Yvette, France\\
	}
	\vspace{1cm}
\end{center}

\vspace{4mm}

\begin{abstract}
We show that the four-point functions in conformal field theory are defined as distributions on the boundary of the region of convergence of the conformal block expansion. The conformal block expansion converges in the sense of distributions on this boundary, i.e.~it can be integrated term by term against appropriate test functions. This can be interpreted as a giving a new class of functionals that satisfy the swapping property when applied to the crossing equation, and we comment on the relation of our construction to other types of functionals. Our language is useful in all considerations involving the boundary of the region of convergence, e.g.~for deriving the dispersion relations.
We establish our results by elementary methods, relying only on 
crossing symmetry and the standard convergence properties of the conformal block expansion. This is the first in a series of papers on distributional properties of correlation
functions in conformal field theory.
\end{abstract}

\vspace{.2in}
\vspace{.3in}
\hspace{0.7cm} January 2019

\newpage

\setcounter{tocdepth}{3}

{
	\tableofcontents
}

\section{Introduction}

Historically, distributions played a big role in axiomatic approaches to
quantum field theory (QFT), via Wightman axioms~\cite{Streater:1989vi} or Osterwalder-Schrader
axioms~\cite{osterwalder1973,osterwalder1975}. In particular, the language of tempered distributions allows clean
treatment of correlation functions singularities at $x^2 = 0$ in a UV-complete
QFT, where $x^2$ may be Euclidean or Lorentzian distance.

In recent years, a new axiomatic approach---the conformal bootstrap---has
emerged in the study of conformal field theories (CFTs) in dimension $d
\geqslant 2$, i.e.\ quantum field theories invariant under the action of
conformal group (see review~\cite{RMP}). This approach is both rigorous and calculable. On the
numerical side, it has allowed precise determinations of many experimentally
measurable quantities, such as the critical exponents of the 3d Ising \cite{ElShowk:2012ht,El-Showk:2014dwa, Kos:2014bka,Simmons-Duffin:2015qma,Kos:2016ysd}, $O (N)$ \cite{Kos:2013tga,Kos:2015mba, Kos:2016ysd, Chester:2019ifh}
and other critical points. On the analytic side, it also led to many insights
into the structure of operator spectrum of general CFTs, in particular
concerning how operators organize themselves in infinite families (Regge
trajectories)~\cite{Komargodski:2012ek,Fitzpatrick:2012yx,Caron-Huot:2017vep}. Numerical bootstrap studies typically take place deep in the
Euclidean region, staying away from the contact term singularities of
correlation functions at short distances. In this regime, the rules of the game
are well-understood and comprise the Euclidean bootstrap axioms. 

On the other hand, analytical
bootstrap studies often boldly go into the Lorentzian space, probe light-cone or other types of singularities. In this regime the most common set of assumptions for correlation functions are the Wightman axioms~\cite{Streater:1989vi}, but it has never been shown how these assumptions follow from the
well-understood Euclidean bootstrap axioms. To achieve this is the goal of
this series of papers. The uniting theme of this work will be tempered
distributions, hence the title.

In this first paper of the series we will study convergence of the conformal
block decomposition. As is well known, it converges in the sense of functions
inside the unit disk $| \rho |, | \bar{\rho} | < 1$ for the radial variable.
We will show that it converges in the sense of distributions also on the
boundary of this unit disk. This is done using Vladimirov's theorem~\cite{Vladimirov}---a key
result in the theory of functions of several complex variables that we will
carefully introduce.

Vladimirov's theorem provides the answer to the following question: if we have a function
$g(\rho)$ that is holomorphic in the open unit disc $|\rho|<1$, what can we say about
its values for $|\rho|=1$? If $g(\rho)$ were bounded, then the limit $\lim_{r\to 1} g(re^{i\theta})$
would be guaranteed to exist for almost every $\theta$ and give rise to a bounded function $g(e^{i\theta})$. However, the functions of cross-ratios that
we encounter in conformal field theory are not bounded and instead can blow up near the boundary.
Crucially though, it is easy to show (as we do in this paper) that they blow up only as power laws
$(1-r)^{-K}$. In this case, Vladimirov's theorem guarantees that the limit $\lim_{r\to 1} g(re^{i\theta})$
exists in the space of distributions in the variable $\theta$. We will explain that this conclusion
holds both for the correlation function itself as well as for the individual terms in the conformal 
block expansion, which will allow us to prove convergence of conformal block expansion in the space of distributions. A simple yet illustrative example of distributional convergence is the sum
\be
	\sum_{n=-\oo}^{+\oo} e^{in\theta}=2\pi \de(\theta),
\ee
where $\theta\in(-\pi,\pi]$ is the coordinate on the unit circle. This sum doesn't converge
in the usual sense because every term is of absolute value 1, but it does converge after being 
smeared with a smooth test function $f(\theta)$. We will study a more realistic toy example in section~\ref{sec:toy}.

Our results can be interpreted as introducing a new class of functionals which satisfy the swapping property~\cite{Rychkov:2017tpc}
when applied to the crossing equation. This point of view might be helpful for readers with interest in analytic functional bootstrap~\cite{Mazac:2016qev,Mazac:2018mdx,Mazac:2018ycv,Kaviraj:2018tfd,Mazac:2018biw,Hartman:2019pcd,Paulos:2019gtx,Mazac:2019shk}.
Specifically, we show that integration (appropriately defined) of the crossing equation with a test function over the boundary of the crossing region\footnote{By crossing region we mean the region in cross-ratio space where both $s$- and $t$-channel conformal block expansions converge. In the standard $z$-cross ratio it is given by $\C$ minus the cuts along $[1,+\oo)$ and $(-\oo,0]$.} can be exchanged
with the sum over conformal blocks. We prove this result for infinitely smooth test functions, and argue that it likely can be strengthened to enlarge the class
of test functions sufficiently so that our new class of functionals will include all functionals currently known to satisfy the swapping property.

In our second paper {\cite{paper2}}, CFT Wightman four-point functions in Lorentzian
space will be shown to be tempered distributions, thus establishing Wightman
axioms. In the third paper {\cite{paper3}}, we will study analytic
continuations of CFT correlation functions to the Lorentzian cylinder (also
known as the boundary of the AdS space). Our goal is to establish everything
from Euclidean bootstrap axioms, without any extra assumptions. When the time
comes, we will explain that the existing classic results in the literature,
like the Osterwalder-Schrader theorem~\cite{osterwalder1973,osterwalder1975} or the construction of L{\"u}scher and
Mack~\cite{Luscher:1974ez}, all require additional assumptions. So our conclusions cannot be recovered from the classic papers.
Fortunately, we found a different way of reasoning which recovers all the results commonly assumed to be true, for the most important in applications case of four-point functions.\footnote{It's an interesting open problem how to extend our arguments to higher point functions.} The good news is that our alternative arguments are really easy, and the main idea can really be summarized in one sentence: ``Look for a powerlaw bound.'' 
This should be contrasted with the classic papers which are quite intricate.

The present paper is organized as follows. In section~\ref{sec:cbexpansion} we discuss the motivation for our work from the 
point of view of computing Euclidean and Lorentzian correlation functions. In section~\ref{sec:1dcase} we consider the simplified
case of one cross-ratio, starting with a toy example of power series. We also use this simplified setting to discuss possible applications
of our results to analytic functional bootstrap (section~\ref{sec:functionals}) and to proper definition of discontinuities (section~\ref{sec:dispersion}).
In section~\ref{sec:higherd} we consider the case of two cross-ratios in scalar correlators in general number of dimensions. We comment on
applications and generalization to spinning correlators. 
In section \ref{sec:Bissi} we discuss an application in the context of a single-variable dispersion relation recently proposed by Bissi, Dey and Hansen \cite{Bissi:2019kkx}. We conclude in section~\ref{sec:conclusions}.

\section{Conformal block expansion}
\label{sec:cbexpansion}

In this section we will state our basic problem, and the main idea how to solve it.
Let us consider the conformal block expansion of a four-point function of identical scalar operators 
(we will consider more general four-point functions later)
\begin{equation}
\label{eq:4pt}
	\<\f(x_1)\f(x_2)\f(x_3)\f(x_4)\>=\frac{1}{(x_{12}^2)^{\De_\f}(x_{34}^2)^{\De_\f}}g(u,v),
\end{equation}
where, as usual
\begin{equation}
	u=\frac{x_{12}^2x_{34}^2}{x_{13}^2x_{24}^2},\quad v=\frac{x_{14}^2x_{23}^2}{x_{13}^2x_{24}^2}.
\end{equation}
We will mostly be working with the radial coordinates $\rho,\bar\rho$~\cite{Pappadopulo:2012jk,Hogervorst:2013sma} defined as
\begin{equation}
\label{eq:rho}
	\rho=\frac{z}{(1+\sqrt{1-z})^2},\quad 	\bar\rho=\frac{\bar z}{(1+\sqrt{1-\bar z})^2},
\end{equation}
where $z,\bar z$ are determined by
\begin{equation}
\label{eq:zzbar0}
	z\bar z=u, \quad (1-z)(1-\bar z)=v.
\end{equation}
We will abuse the notation a bit by writing $g(u,v), g(z,\bar z)$, or $g(\rho,\bar\rho)$ depending
on which set of cross-ratios we want to use.

The function $g(\rho,\bar \rho)$ can be expanded in conformal blocks in $\f(x_1)\times\f(x_2)$ OPE channel as follows,
\be
\label{eq:cbexpansion}
g(\rho,\bar \rho) = \sum_{\De,J} p_{\De,J} g_{\De,J} (\rho,\bar \rho),
\ee
where $p_{\De,J}\geq 0$ are the OPE coefficients squared, and $g_{\De,J}(\rho,\bar \rho)$ 
are the conformal blocks. This expansion is known to be absolutely convergent in the region $|\rho|<1,|\bar\rho|<1$, which we will denote by $\cC$ in what follows.

We will only use the global conformal invariance $SO(d+1,1)$. Under these assumptions, the region $\cC$ is the largest region of convergence of the conformal block decomposition of a general CFT four-point function (we are not aware of any results to the contrary). In 2d CFT, using Virasoro, the region of convergence can be extended further in terms of Al.~Zamolodchikov's uniformizing $q$ variable, being given by $|q|,|\bar q|<1$ which is a strictly larger region than $\cC$~\cite{Hogervorst:2013sma,Maldacena:2015iua}. So our results should be best possible in $d>2$ but not in $d=2$.

Above we focused on the 12 OPE channel ($s$-channel) but the same discussion can be made for the $t$-channel 23 and $u$-channel 13, whose convergence is characterized by the conditions $|\rho_t|,|\bar\rho_t|<1$ and $|\rho_u|,|\bar\rho_u|<1$.

Let us briefly describe what the region $\cC$ corresponds to in the physical space of $x_i$.
In Euclidean signature, this region includes all configurations when the four points $x_i$ do not lie on a circle, which is the generic case. 
If $x_i$ do lie on a circle, the cross-ratios belong to $\cC$ if $x_1$ and $x_2$ are next to each other on the circle. If the points instead fall on the 
circle in the ordering $x_1, x_3, x_2, x_4$ (read in some direction), then we find $|\rho|=|\bar\rho|=1$. Therefore, only a measure zero set of Euclidean configurations does not 
belong to $\cC$ and is instead on its boundary $\ptl \cC$. For these configurations the $s$-channel expansion does not converge. However, it does converge for $t$- and $u$-channels, and so
the value of the Euclidean four-point function can be determined from the OPE for any configuration of the four points.

Our basic problem is to make sense of the four-point function~\eqref{eq:4pt} in Lorentzian signature. 
In order to talk about a Lorentzian four-point function, we need to specify which operator ordering we are interested in. We will only consider here the Wightman functions, i.e.~we fix operator ordering:
\begin{equation}
\label{eq:bestordering}
W(x_1,x_2,x_3,x_4)= \<0|\f(x_1)\f(x_2)\f(x_3)\f(x_4)|0\>\,.
\end{equation}
However, we wish to consider all possible time and causal ordering of the points $x_i$.\footnote{Other often considered Lorentzian correlators (retarded, advanced, time-ordered) can be obtained by multiplying Wightman functions with appropriate factors enforcing the needed ordering. The Wightman functions being distributions, and the time-ordering factors being singular, this procedure introduces extra singularities and requires care in a rigorous treatment.} Once we have fixed the operator ordering, the Lorentzian
four-point function can be obtained from the Euclidean one by an appropriate analytic continuation. While in Euclidean we always have
$\bar\rho=\rho^*$, this property is generally lost after the analytic continuation. Furthermore, there are open regions in the Lorentzian configuration space of $x_i$ where $|\rho|$ and/or $|\bar \rho|$ end up $\ge 1$ after the analytic continuation. Then the corresponding conformal block expansion~\eqref{eq:cbexpansion} diverges and thus cannot be used to determine the correlator.

\begin{figure}[h!]
		\centering
		\includegraphics[scale=0.8]{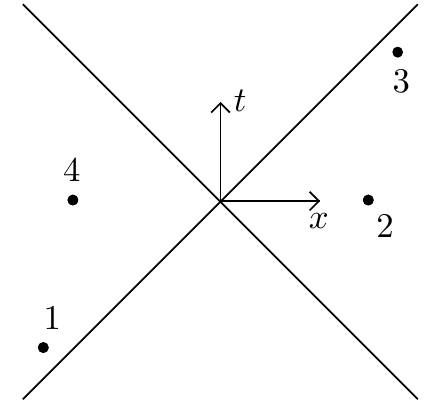}
		\caption{\label{fig-regge} Regge kinematics.}
	\end{figure}

One such well known case is the Regge regime~\cite{Cornalba:2007fs,Cornalba:2008qf,Costa:2012cb}, when $x_1,x_4$ and $x_2,x_3$ pairs are timelike separated, while all other intervals are spacelike (see Fig.~\ref{fig-regge}).
One may be tempted to use the 13 OPE for this Lorentzian correlator, because this channel is the most symmetric with respect to the origin, and also because one may be interested in the limit $x_{13}^2\to 0$. However, although this channel converges when points $x_1$ and $x_3$ stay close to the origin, it starts diverging when they cross the lightcones of $x_2$ and $x_4$ and move into the Regge regime, because the corresponding $\rho_u$ variable become larger than 1.\footnote{Conformal Regge theory~\cite{Cornalba:2007fs,Cornalba:2008qf,Costa:2012cb} provides a way to resum the expansion~\eqref{eq:cbexpansion} in the limit $\bar\rho_u\to0$, $\rho_u\to\oo$ with $\rho_u\bar\rho_u$ fixed. We will not consider such resummations in this paper since they rely on analytically-continued OPE data that we have little control over.} In this particular case, one can switch to the 23 OPE for which $|\rho_t|,|\bar \rho_t|<1$ is less than 1, and so this channel converges. However, this is not always possible: there exist kinematic configurations when no channel converges (appendix~\ref{app:lorentz}).
 
In this series of works we will propose a different way to solve this problem, and recover the Wightman function in all kinematic configurations. In our construction the key role will be played by the 12 OPE-channel. We call it the ``vacuum channel'', because it involves the two leftmost operators  in the Wightman ordering~\eqref{eq:bestordering}, i.e.~the ones acting on the vacuum. While the vacuum channel OPE does not always converge, it \emph{almost converges} for all possible configurations. What this means is that $|\rho|\leq 1$ and $|\bar\rho|\leq 1$ for all values of $x_i$. This crucial fact will be shown in~\cite{paper2}. It is only true for the vacuum OPE channel, but would not be true for the 23 or 13 channels, for which sometimes $|\rho|$ and/or $|\bar\rho|$ will be strictly greater than 1. In particular, as we show in appendix \ref{app:lorentz}, there exist configurations for which both 23 and 13 channels diverge with $|\rho|,|\bar \rho|>1$, while for 12 channel $|\rho|=|\bar \rho|=1$.


In other words, all possible Lorentzian configurations belong to the closure $\bar \cC$. One can ask how large are the regions in configuration space of $x_i$ which belong in $\ptl \cC$ but not in $\cC$.
In Euclidean, we have seen that these configurations were measure zero, but in Lorentzian this is no longer true: extended regions with non-empty interior have $|\rho|=1$, $|\bar\rho|=1$. So, a fraction of configurations are in $\ptl \cC$ and not in $\cC$. 
 
If the conformal block expansion converged in $\bar \cC$ and not $\cC$, we would be able to use it to compute any Lorentzian correlator
in any configuration of the points $x_i$. Of course, this is not the case, and the conformal block expansion converges in the usual sense only in $\cC$. However, our goal in this paper
will be to extend the notion of convergence so that it will become valid in $\bar\cC$. Specifically, we will show that the expansion~\eqref{eq:cbexpansion}
converges in the sense (to be clarified below) of distributions on the boundary $\ptl\cC$ in the cross-ratio space. In the forthcoming work~\cite{paper2,paper3}
we will extend this result to convergence in the sense of distributions in the physical space of $x_i$, either in Minkowski space, or on the Lorentzian cylinder.

One may be wondering what is special about the vacuum channel compared to other OPE channels. Intuitively, the distinguishing feature of vacuum channel is that we can understand it as inserting a complete set of states in the Wightman four-point function. Since Wightman four-point functions are distributions, we cannot generally expect this sum to make sense in terms of functions, but only in terms of distributions. Mack~\cite{Mack:1976pa} understood the vacuum channel OPE expansion in distributional sense in position space. Mack's reasoning is rather nontrivial, and it crucially relies on assuming from the start that Wightman axioms hold in Lorentzian signature---an assumption that we are here not willing to accept. Although our results in cross-ratio space are inspired by Mack's considerations in position space, they do not follow from his results, since we rely on a different and simpler set of assumptions, natural from the modern bootstrap perspective. Also, we are only using rather elementary methods. Position space will be discussed in~\cite{paper2}.

\section{One-dimensional case}
\label{sec:1dcase}

First, let us simplify the problem by considering the one-dimensional case where there is a single cross-ratio. The conformal block expansion takes the form
\begin{equation}
\label{eq:1dcbexpansion}
	g(\rho) = \sum_{\De} p_\De g_\De (\rho),
\end{equation}
where the conformal blocks are given by\footnote{This equation follows from the more familiar one in the $z$ coordinate $g_\De(z)=z^\De {}_2 F_{1}(\De,\De;2\De;z)$ by a hypergeometric identity~\cite{Hogervorst:2013sma}.}
\begin{equation}
\label{eq:1dblock}
	g_\De(\rho) = (4\rho)^\Delta {}_2 F_{1}(1/2,\De;\De+1/2;\rho^2).
\end{equation}
Furthermore, the sum is over $\De\geq 0$ and $p_\De\geq 0$. This expansion converges, in the usual sense, in the interior of the unit disk $|\rho|<1$, and our goal is to understand whether it can be made convergent, in some generalized sense, on the boundary $|\rho|=1$. 

If we look at~\eqref{eq:1dcbexpansion} more closely, we will notice that the conformal blocks~\eqref{eq:1dblock} are not single-valued in the unit disk of $\rho$. Therefore, we
are really interested in the behavior of this sum on the universal cover of the unit disk branched at $0$, which can be conveniently parametrized by writing
\begin{equation}
	\rho=e^{i\tau}.
\end{equation}
The expansion~\eqref{eq:1dcbexpansion} is then absolutely convergent in the upper-half plane of $\tau$, and we are interested in its convergence for real $\tau$.

	\begin{figure}[h!]
	\centering
	\includegraphics[scale=0.7]{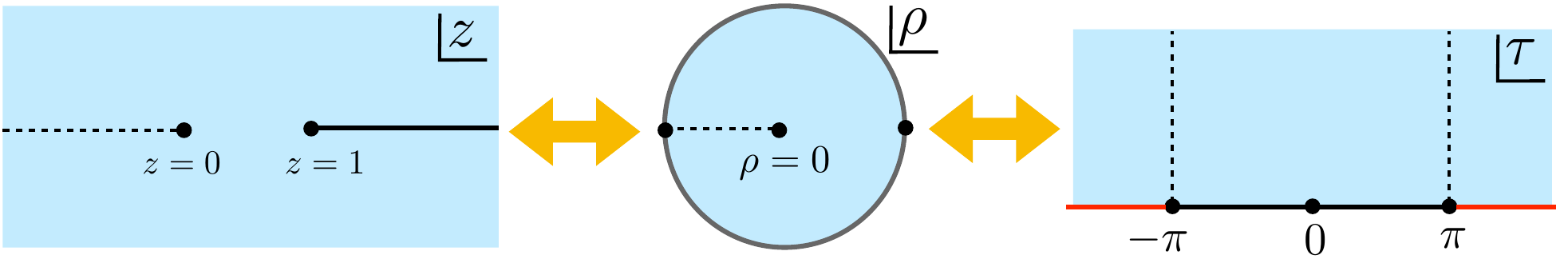}
	\caption{\label{fig:zrhotau} Transformation from the $z$ cut plane to the $\rho$ disk to the $\tau$ upper-half plane, see the text.}
\end{figure}

In Fig.~\ref{fig:zrhotau} we show the transformation from the $z$ cut plane to the $\rho$ disk to the $\tau$ upper-half plane. The two sides of the cut $z\in[1,+\infty)$ are mapped on the boundary of the unit disk $|\rho|=1$, and then to the black part ($\tau\in[-\pi,\pi]$) of the upper-half plane boundary. The rest of the $\tau$ boundary (marked in red) can be accessed in the $\rho$ variable by first going through the cut $\rho\in[-1,0]$ (dashed) and then approaching $|\rho|=1$. 

On the black interval $\tau\in[-\pi,\pi]$ (except at $\tau=0$) the four-point function is actually analytic, as can be shown using the $t$-channel expansion. On the rest of the boundary (red part), the $t$-channel expansion does not converge and provides no information. Below we will show, using the $s$-channel, that the four-point function is a tempered distribution on the whole boundary. We will also show that the $s$-channel conformal block expansion converges in the sense of distributions. When using the $s$-channel, we have to use distributional convergence even on the black part of the boundary, although the function itself is analytic there as explained above. 

\subsection{A toy problem}
\label{sec:toy}

In order to gain some intuition, it is useful to consider the following toy problem. Let us study the power series
\begin{equation}
	\frac{1}{1-\rho}=\sum_{n=0}^\oo \rho^n.
\end{equation}
It has the similar feature that it converges absolutely for $|\rho|<1$ and that the resulting function has a power-like singularity at $\rho=1$, much like the physical four-point functions do.

In terms of $\tau$ variable we find the sum
\begin{equation}\label{eq:tausum}
	\sum_{n=0}^\oo e^{in\tau},
\end{equation}
which clearly does not converge for any real $\tau$. We claim that it does converge as a tempered distribution.\footnote{A tempered distribution is a distribution that can be paired with Schwartz test functions (see below).} For example, let us compute its real part using the standard formulas of Fourier analysis
\begin{equation}
	\mathrm{Re}\sum_{n=0}^\oo e^{in\tau}=\half\sum_{n=-\oo}^\oo e^{in\tau}+\half=\half+\pi \sum_{k=-\oo}^\oo \de(\tau-2\pi k).
\end{equation}
It is a bit harder to compute the imaginary part, but we can run the following simple argument for the full sum~\eqref{eq:tausum}. Let $f(\tau)$ be a Schwartz test function, i.e. a smooth\footnote{In this paper ``smooth'' means $C^\infty$ and the two terms are used interchangeably.} function which, together with its derivatives, decays at infinity faster than any power. In order
to show that~\eqref{eq:tausum} converges as a tempered distribution, we need to show, by definition, that the partial sums
\begin{equation}\label{eq:toypartialsums}
	 \int d\tau f(\tau) \sum_{n=0}^N e^{in\tau}= \sum_{n=0}^N \int d\tau f(\tau) e^{in\tau}=\sum_{n=0}^N \tl f(n)
\end{equation}
converge to a finite limit as $N\to \oo$. Here, $\tl f(n)$ is the Fourier transform of $f$. Since $f(\tau)$ is a Schwartz test function, so is $\tl f(n)$ (where $n$ is understood as a real parameter) and thus
$\tl f(n)$ decays faster than any power of $n$ as $n\to \oo$. This implies that the partial sums~\eqref{eq:toypartialsums} indeed converge. Strictly speaking, we also need to show that the limit is continuous with respect to $f$ in an appropriate topology. We will delay this question until later. Here the important message is that even though~\eqref{eq:tausum} does not converge in the usual sense, it starts to converge after being smeared with a nice test function.

So far we have learned two things. First, the sum~\eqref{eq:tausum} converges in distributional sense for real $\tau$. Second, the value of this sum is a genuine distribution, since we computed its real part and it is a sum of $\de$-functions. Now, we also know that in the upper-half plane of $\tau$ the sum converges to 
\begin{equation}
	g(\tau)=\frac{1}{1-\rho}=\frac{1}{1-e^{i\tau}}.
\end{equation}
This suggests that on the real line $g(\tau)$ should have a limit that is the tempered distribution computed by~\eqref{eq:tausum}. So we can conjecture that, for real $\tau$,
\begin{equation}
\label{eq:conjecture}
	\sum_{n=0}^\oo e^{in\tau} = \lim_{\epsilon \to +0} g(\tau+i\e)\equiv \lim_{\e\to +0}\frac{1}{1-e^{-\e+i\tau}},
\end{equation}
where everything is understood in the sense of tempered distributions. 

How can we guarantee that the limit in the right-hand side exists? In the sense of functions, it clearly exists for $\tau\neq 2\pi k$ and is given by $g(\tau)$. However, $g(\tau)$ for real $\tau$ is not obviously a distribution, since it involves non-integrable singularities near $\tau=2\pi k$ that we need to regulate. Specifically, we need to prove that for any Schwartz function $f(\tau)$ the limit
\begin{equation}
\label{eq:limit-toy}
	\lim_{\epsilon \to +0} \int d\tau g(\tau+i\e) f(\tau)
\end{equation}
exists and depends continuously on $f$ in an appropriate topology. Notice that if $f(\tau)$ were a holomorphic function, for example $f(\tau)=e^{-\tau^2}$, then the existence of the limit would be simple to show.\footnote{We will just give an idea. Expand $f(\tau)$ in Taylor series around $f(\tau+i\eps)$ as $f(\tau)=f(\tau+i\eps)+ (-i\eps) f'(\tau+i\eps)+\ldots+O(\eps^m)$. The terms involving $f^{(k)}(\tau+\eps)$ are easy to analyze: the integrals don't depend on $\eps$ at all because by analyticity we can shift the contour. So only the first of these terms survives. The error term goes to zero provided that $\eps^m g(\tau+i\eps)\to 0$ as $\eps\to0$. This will hold for $m>M$ if $g$ satisfies the slow-growth condition~\eqref{eq:slow-growth} below. This shows that one could equivalently define the pairing between $g$ and holomorphic $f$ by shifting the integration contour for both, as $\int d\tau g(\tau+i\e) f(\tau+i\e)$. This is independent of $\e$ and there is no limit to talk about.} However, the class of holomorphic test functions is too restricted for many purposes. It is more customary to develop the theory of distributions using compactly supported $C^\infty$ test functions, or the even larger class of Schwartz test functions.\footnote{This is not just for the reasons of generality. Compactly supported test functions are needed if one wants to define a very basic notion of support of the distribution. This notion allows as to make statements such as ``distributions $f(x)$ and $g(x)$ agree for $x\in [0,1]$ but disagree outside of this interval.'' Then the class of Schwartz test functions, being invariant under the Fourier transform, plays an important role in all questions involving the Fourier transform of distributions.} For a general Schwartz $f(\tau)$, existence of the limit~\eqref{eq:limit-toy} requires an argument which will be explained in the next section.

\begin{figure}[h!]
	\centering
	\includegraphics[scale=0.5]{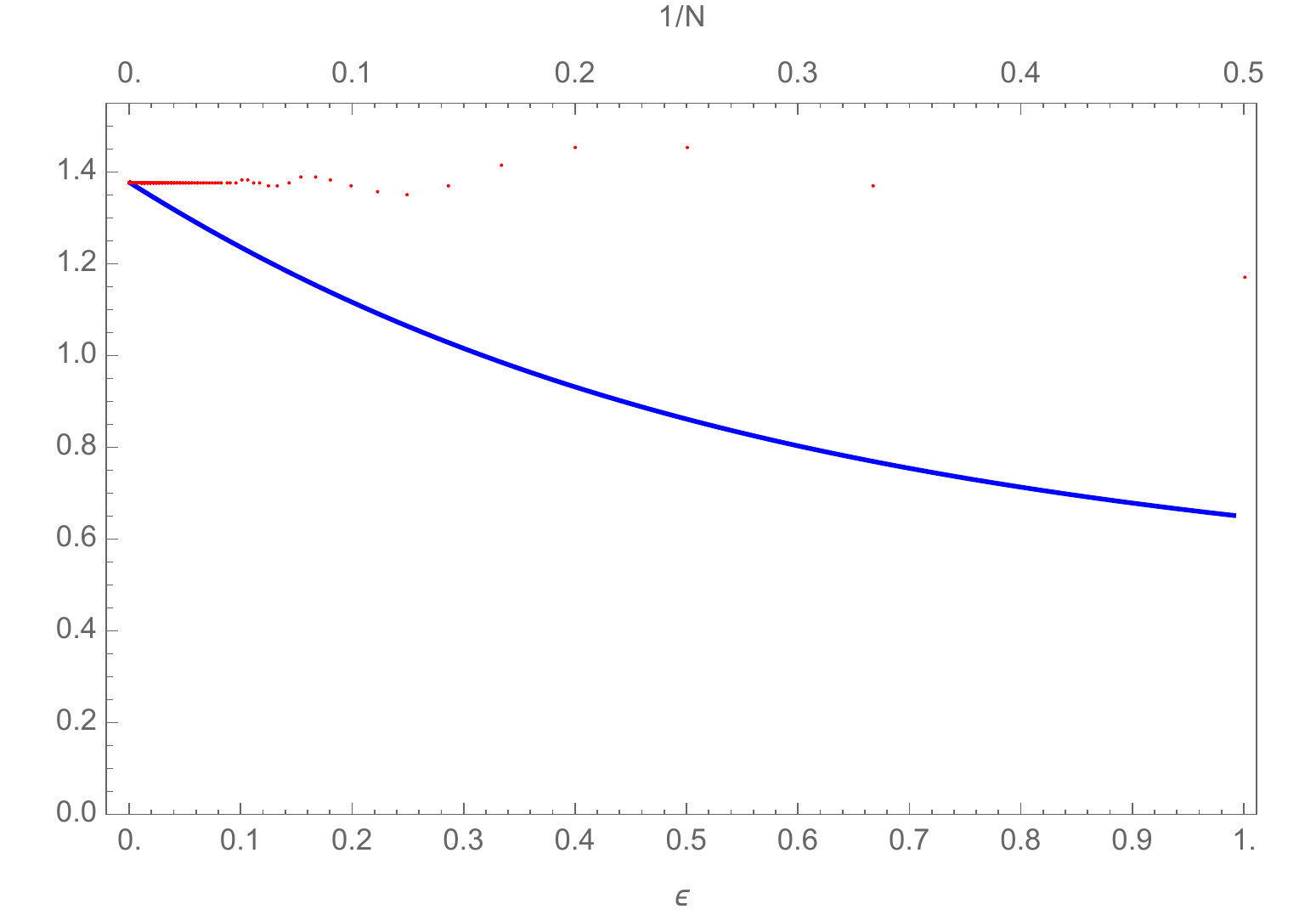}
	\caption{\label{fig-limit} A numerical check of the existence of the limit~\eqref{eq:test-limit}, for $f(\tau)$ given in the text. The curve is the integral under the limit sign, and the red dots are the partials sums of Fourier coefficients in the r.h.s.\ of~\eqref{eq:test-limit} up to $n=N$.}
\end{figure}

We would like to emphasize that the existence of the limits~\eqref{eq:conjecture},~\eqref{eq:limit-toy} is not just some abstract nonsense, but a very concrete prediction. Integrating both parts of~\eqref{eq:conjecture} against an arbitrary Schwartz test function $f(\tau)$, we obtain:
\beq
\label{eq:test-limit}
\lim_{\e\to +0}\int d\tau\,f(\tau)\frac{1}{1-e^{-\e+i\tau}} = \sum_{n=0}^\infty \tl f(n)\,.
\eeq
Let us test this prediction. We pick a function $f(\tau)$ given by $\exp(-1/(1-\tau^2))$ for $\tau\in(-1,1)$, extended by zero outside this interval. It is a compactly supported $C^\infty$ function (in particular Schwartz, but not analytic). We evaluate both sides of the previous equation numerically for $0<\e<1$, and check the limit (see Fig.~\ref{fig-limit}).

\subsection{Vladimirov's theorem}
\label{sec:proof1}

Fortunately, there is a general result that immediately establishes that~\eqref{eq:conjecture} is valid, i.e.\ that both left and right hand sides converge as tempered distributions and are indeed equal. Before stating this result, let us first clean up some formal definitions. 

For a smooth function $f(x)$ define the semi-norms
\begin{equation}
\label{eq:semi-norm}
	\|f\|_{m,n}=\sup_{x \in \R} |(1+|x|^m) \ptl_x^n f(x)|.
\end{equation}
The Schwartz space $\cS(\R)$ consists of smooth functions $f$ for which $\|f\|_{m,n}$ is finite for all non-negative integer $m$ and $n$. 
This is a vector space which is given a topology where a sequence $f_k$ is said to converge to $g$ if $h_k=f_k-g$ converges to $0$. In turn,
$h_k$ converges to $0$ iff for all $m,n$ the sequence $\|h_k\|_{m,n}$ converges to $0$.

The space $\cS'(\R)$ of tempered distributions is defined as the space of continuous linear functionals on $\cS(\R)$. We say that
a linear functional $\a$ is continuous if $\a(h_k)\to 0$ for any sequence $h_k\in \cS(\R)$ for which $h_k\to 0$.
We say that a sequence of tempered distributions $\a_k$ converges to a tempered distribution $\beta$ if for any $f\in \cS(\R)$ we have
$\a_k(f)\to \beta(f)$.

Now, let $a>0$ and $g(\tau)$ be a function holomorphic in the strip $0<\mathrm{Im}\,\tau < a$. Suppose there exist $N,M\in \Z_{\geq 0}$ and $C>0$ such that in the strip
\begin{equation}
\label{eq:slow-growth}
	|g(x+iy)|\leq C(1+|x|^N)y^{-M}
\end{equation}
for all $x\in \R$ and $y\in (0,a)$. We then say that $g$ satisfies a slow-growth condition near $\R$. What this means is that for any $y$ the function $g(x+iy)$ is bounded by a polynomial of fixed degree, and the overall size of this polynomial grows at most as a fixed powerlaw when $y\to0$. Note that thanks to this condition for any $y$, $0<y<a$, the function $g_y(x) \equiv g(x+iy)$
is a tempered distribution in $\cS'(\R)$. We can ask whether the limit
$
	\lim_{y\to +0} g_y
$
exists in $\cS'(\R)$. If it does, we say that boundary value of $g$ on $\R$ exists in $\cS'(\R)$ and denote it by $\mathrm{bv}\, g$,
\begin{equation}
	\mathrm{bv}\, g \equiv \lim_{y\to +0} g_y.
\end{equation}

We can now state the theorem
\begin{theorem}
	\label{thm:vlad}
	Let $g(\tau)$ be a function holomorphic for $0<\mathrm{Im}\,\tau<a$ for some $a>0$, satisfying the slow-growth condition near $\R$ as defined above.
	Then the boundary value $\mathrm{bv}\,g$ of $g$ on $\R$ exists in $\cS'(\R)$. Furthermore, if a sequence of functions $g_n$, holomorphic in the same region, satisfies the slow-growth condition with the same constants $C,M,N$ for all $n$ (uniform slow-growth condition), and 
	converges pointwise to $g$ for $0<\mathrm{Im}\,\tau<a$, then 
	$g$ satisfies the same slow-growth condition and
	\begin{equation}
		\lim_{n\to \oo} \mathrm{bv}\, g_n = \mathrm{bv}\, g\quad\text{in }\cS'(\R).
	\end{equation}
\end{theorem}

Such results are rather standard in the theory of distributions (an early mathematics reference is~\cite{Tillmann1961}). In mathematical physics they are very useful in the study of QFT Wightman functions. The standard reference is the book of Vladimirov~\cite{Vladimirov} (section 26), and we will therefore refer to such results as ``Vladimirov's theorems''. A self-contained proof of Theorem~\ref{thm:vlad} will be given below. A more general Vladimirov's theorem will be stated and used in~\cite{paper2}. 

Let us see how this result applies to our toy problem. We have
\begin{equation}
	g_n(\tau)=\sum_{k=0}^n e^{ik\tau}, \quad g(\tau)=\frac{1}{1-e^{i\tau}}.
\end{equation}
Let us check the slow growth condition for $g_n$ on $0<\mathrm{Im}\,\tau <1$:
\begin{equation}
	|g_n(x+iy)|\leq \sum_{k=0}^n |e^{ikx-ky}|\leq \sum_{k=0}^\oo e^{-ky}=\frac{1}{1-e^{-y}}\leq Cy^{-1}
\end{equation}
for some $C>0$. So we see that the slow growth condition is satisfied with $N=0, M=1$. The same condition is then
true for $g(\tau)$, as is easy to check. Then theorem~\ref{thm:vlad} immediately implies our conjecture~\eqref{eq:conjecture}.

\subsection{Proof of Vladimirov's theorem~\ref{thm:vlad}}
\def\L{L}
\label{sec:proof}

We first prove that $\mathrm{bv}\,g$  exists and is a tempered distribution. So we pick a Schwartz test function $f(x)$ and study the integral
\begin{equation}
\L(y):= \int dx\,  g (x + i y) f (x)   . 
\label{hgen}
\end{equation}
We need to show that this has a limit as $y \rightarrow +0$. This looks a bit
magic: estimating naively by absolute value one would conclude that
the integral may blow up as $y^{-M}$. It won't blow up only because of
cancellations, not captured by the naive estimate. In other words, when an analytic function tends somewhere to infinity, it will tend to minus infinity nearby, so that the integral will remain finite. For intuition, recall the
Sochocki formula:
\be 
\lim_{y \rightarrow +0} \overline{} \frac{1}{x + i y}
= \text{PV} \frac{1}{x} - i \pi \delta (x)  
\ee
Principal value PV represents a kind of cancellations whose existence we need to exhibit in
general.

Going back to {\eqref{hgen}},\footnote{We follow the proof in
	{\cite{Streater:1989vi}}, Theorem 2-10.} the first key idea is that we can
estimate not just $\L$ but any its derivative. 
By the
Cauchy-Riemann equations, $y$-derivatives of $\L (y)$ can be transformed into
$x$ derivatives acting on $g$ which then can be integrated by parts to act on
$f$:
\be
	L^{(j)}(y)=i^j\int dx g^{(j)}(x+iy) f(x)=(-i)^j \int dx g(x+iy) f^{(j)}(x).
\ee 
Using then the slow-growth condition {\eqref{eq:slow-growth}} we get an estimate of
\emph{any} $y$-derivative $\L^{(j)} (y)$ by $y^{-M}$ times a constant:
\be
\label{eq:jbound}
| L^{(j)} (y) | \leqslant C y^{-M}\,.
 \ee
The constant here is proportional to the semi-norm $\|f\|_{N+2,j}$, see~\eqref{eq:semi-norm}; order $N+2$ is needed to make the integral convergent, while derivative order $j$ appears because of integrating by parts. 

This is still growing as $y \rightarrow 0$. Here comes the second key idea:
since we have this bound on any derivative, we can strengthen it recursively
using the Newton-Leibnitz formula:
\be \label{eq:NLformula}
\L^{(j - 1)} (y) = - \int_{y^{}}^{y_0} dy\,\L^{(j)} (y) + \L^{(j - 1)} (y_0) \,.
\ee
Here $y_0$ can be any fixed number in the strip of analyticity, e.g. $y_0=a/2$ will do.

Every time we use this, we obtain a bound on $\L^{(j - 1)}$ of the same type as in~\eqref{eq:jbound} but with the order of
singularity in $y$ reduced by 1 w.r.t.~$\L^{(j)}$. Let us do this repeatedly, starting from $j=M+2$.\footnote{Exercise: once you understand the proof below, show that $j=M+1$ will do as well. Hint: the key requirement is that $\L'(y)$ end up bounded by some integrable function.} Then doing this $M$ times we will prove that $\L''(y)$ has an at most $\log(y)$ singularity, and doing this once more we prove that $\L'(y)$ has no singularity at all, i.e.~it is bounded by a constant, call it $C_1$.

Now we can finally prove that $\L(y)$ has a limit. From the $j=1$ case of~\eqref{eq:NLformula} we can write
\beq
(\text{bv}\, g)(f) = \lim_{y\to+0} \L(y)=-\int_{0}^{y_0} dy\,\L'(y)+L(y_0)\,.
\eeq
The limit exists, since by $|\L'(y)|\le C_1$ the integral in the r.h.s.~converges absolutely at the lower limit of integration.
Thus $\text{bv} \,g$ exists as a linear functional on $\cS(\mathbb{R})$. All constants in the above argument are bounded by some semi-norms of $f$. This proves that $\text{bv}\,g$ is a \emph{continuous} linear functional on $\cS(\mathbb{R})$, i.e.~a tempered distribution.

%
%

Now let us prove the second part of the theorem, about convergence. Replacing $g_n$ by $g_n-g$, it's enough to consider the case $g=0$. We pick an arbitrary Schwartz function $f$ and consider
\beq
\label{eq:bvgn}
(\text{bv}\,g_n)(f)=\lim_{y\to +0} L_n(y)\,.
\eeq
Here $L_n(y)$ is defined by the integral~\eqref{hgen} with $g$ replaced by $g_n$. The existence of the limit for each $n$ is guaranteed by the above argument. As a byproduct of the argument, we have also seen that  $|L'_n(y)|\le C_1$ uniformly in $n$ and $y$, where $C_1$ is bounded by some semi-norm of $f$. 

Furthermore, we claim that $L_n(y)$ tends to zero as $n\to\infty$ for any fixed $y\in(0,a)$. Indeed the integrand in~\eqref{hgen} satisfies two conditions: (a) it tends to zero as $n\to \infty$ because $g_n(x+iy)$ goes pointwise to zero; (b) it is bounded in absolute value by an integrable function which does not depend on $n$:
\beq
|g_n(x+iy)f(x)|\leq \|f\|_{N+2,0}\frac{|g_n(x+iy)|}{1+|x|^{N+2}} \le C\|f\|_{N+2,0}\frac{1+|x|^N}{y^M(1+|x|^{N+2})}\,,
\eeq
where we bounded $f$ by its semi-norm, and then used the slow-growth condition~\eqref{eq:slow-growth}.
So the claim follows by Lebesgue's dominated convergence theorem.

Finally we wish to prove that $(\text{bv}\,g_n)(f)$ tends to zero as $n\to\infty$, as this is what is meant by $\text{bv}\, g_n\to 0$ in $\cS'(\mathbb{R})$. From definition~\eqref{eq:bvgn}, we can bound this quantity as:
\be
|(\text{bv}\,g_n)(f)|\le \sup_{y\in(0,\eps)} | L_n(y)| \le |L_n(\eps)|+C_1\eps\,,
\ee
where in the second inequality we used $|L'_n(y)|\le C_1$. We proved above that $L_n(\eps)$ goes to zero for any $\eps$. So by picking first $\eps$ small enough, and then $n$ large enough, the sum of the two terms in the r.h.s.~is arbitrarily small. This implies that $\limsup_{n\to\infty} |(\text{bv}\,g_n)(f)|$ is arbitrarily small. Thus it is zero.

The attentive reader may notice that the last steps of the proof are not constructive, i.e.~they do not provide a bound on how fast $(\text{bv}\,g_n)(f)$ tends to zero. This is because the used assumption, that $g_n$ converges to zero pointwise, is very general. It allows to conclude, via dominated convergence, that $L_n(y)$ tends to zero pointwise as $n\to\infty$, but it does not tell us how fast this limit is reached. If more detailed information about the rate of the limit $g_n\to 0$ is available, as it usually is in practical applications, then a simple modification of the above argument makes the conclusion $\text{bv}\,g_n \to 0$ in $\cS'(\mathbb{R})$ constructive.

\subsection{Distributional convergence of conformal block expansion}
\label{sec:distributionalconvergence1d}


Let us now turn back to the 1-dimensional conformal block expansion~\eqref{eq:1dcbexpansion}. We would like to claim that it converges as a tempered distribution for real $\tau$ (recall $\rho=e^{i\tau}$). To prove this,
we will use Vladimirov's theorem~\ref{thm:vlad}, for which we need to establish a uniform slow-growth condition on the partial sums in the left-hand side of~\eqref{eq:1dcbexpansion}.

As a first step, let us derive a slow-growth condition for the four-point function $g(\rho)$ itself. First, note that for $|\rho|<1$ we have
\be\label{eq:1dscalingexpansion}
	g(\rho)=\sum_\De \tl p_\De \rho^\De.
\ee
with some positive coefficients $\tl p_\De$. This follows from radial quantization in an appropriate conformal frame~\cite{Hogervorst:2013sma}. Equivalently, we can expand the conformal blocks~\eqref{eq:1dblock} in the right-hand side of~\eqref{eq:1dcbexpansion} in powers of $\rho$ and use the fact that these expansions have positive coefficients. In particular, the sum~\eqref{eq:1dscalingexpansion} can be turned back into the sum~\eqref{eq:1dcbexpansion} by appropriately grouping the terms. Now, we can write
\be
	|g(\rho)|=\left|\sum_\De \tl p_\De \rho^\De\right|\leq \sum_\De \tl p_\De |\rho|^\De = g(|\rho|),
\ee
so it suffices to bound $g(\rho)$ for real $\rho\in (0,1)$. This maps to $z\in(0,1)$, and in terms of $z$ variable we know that $g(z)$ satisfies the crossing equation
\be
\label{eq:crossing1d}
	z^{-2\De_\f}g(z)=(1-z)^{-2\De_\f}g(1-z).
\ee
When $z\to 1$, we have $g(1-z)=O(1)$, which implies for $z\in (0,1)$ the bound
\be
	|g(z)|\leq C(1-z)^{-2\De_\f}
\ee
for some $C>0$. Using the fact that $1-z\sim (1-\rho)^2/4$ as $z\to 1$, we find
\be
\label{eq:b1}
	|g(\rho)|\leq g(|\rho|) \leq C' (1-|\rho|)^{-4\De_\f}
\ee
for some $C'>0$. In terms of $\tau = x+iy$ this implies a powerlaw bound
\be
\label{eq:b2}
	|g(\tau)|\leq C''y^{-4\De_\f},
\ee
near $y=0$ for a $C''>0$, which is the required slow-growth condition. Therefore, by theorem~\ref{thm:vlad}, $\text{bv}\, g$ exists and is a tempered distribution.

An easy modification establishes the slow-growth condition for the partial sums in~\eqref{eq:1dcbexpansion}. Let $I$ be any (possibly infinite) subset of the terms in~\eqref{eq:1dscalingexpansion} and write
\be
\label{eq:b3}
	\left|\sum_{\De\in I}\tl p_\De \rho^\De\right| \leq \sum_{\De\in I}\tl p_\De |\rho|^\De \leq \sum_\De \tl p_\De  |\rho|^\De=g(|\rho|)\leq C''y^{-4\De_\f}.
\ee
Taking $I=I_{\De_*}=\{\De|\De<\De_*\}$ we get a uniform slow-growth condition for partial sums of~\eqref{eq:1dscalingexpansion}. Similarly, by allowing $I=I_n$ to contain the terms corresponding to the first $n$ conformal blocks in~\eqref{eq:1dcbexpansion} we get a uniform slow-growth condition on partial sums of~\eqref{eq:1dcbexpansion}. Therefore, by theorem~\ref{thm:vlad}, we conclude that the expansion~\eqref{eq:1dcbexpansion} converges for the boundary values,
\be
	\mathrm{bv}\, g = \sum_{\De} p_\De \mathrm{bv}\, g_\De\quad\text{in }\cS'(\R). 
\ee

Let us unpack this equation a bit. Notice that in the case at hand, $\mathrm{bv}\, g_\De$ is an ordinary locally integrable function which is the easiest kind of distribution. This is because the conformal blocks~\eqref{eq:1dblock} only have a logarithmic singularity at $\rho=1$. Written in full, this equation says that for any Schwartz function $f(\tau)$
\be
\lim_{\eps\to +0} \int d\tau\, g(\rho=e^{-\eps+i\tau}) f(\tau) = \sum_{\De} p_\De \int d\tau\, (\mathrm{bv}\, g_\De)(\rho=e^{i\tau}) f(\tau)\,,
\ee
in the sense that the ${\eps\to +0}$ limit in the l.h.s.~exists (it defines $(\mathrm{bv}\, g)(f)$), the series in the r.h.s.~made of ordinary integrals converges, and that the two sides independently defined in this way are equal.


\subsection{Convergence for other normalizations and on other boundaries}
\label{sec:1dvariants}

We have proven that the conformal block expansion~\eqref{eq:1dcbexpansion} converges as a distribution on the boundary $\ptl \cC$ of the normal function-like domain of convergence $\cC$. We motivated this question in section~\ref{sec:cbexpansion} from the point of view of computing the Wightman functions. However, in other applications the domain $\cC$ may not be the most natural one to consider. For example, 
one of the main objects of study in CFT is the crossing equation 
\be
	\label{eq:1dcrossing}
		z^{-2\De_\f}g(z)=(1-z)^{-2\De_\f}g(1-z)\,,
\ee
where both left- and right-hand side are expanded in conformal blocks. The two expansions are conventionally referred to as the $s$- and $t$-channel expansions. It is then natural to consider the domain $\cC^{st}=\cC^s\cap \cC^t$ in which both expansions converge in the sense of functions, as well as distributional convergence on its boundary $\ptl \cC^{st}$. Additionally, the function $g(z)$ is multiplied by a factor $z^{-2\De_\f}$ in the above equation, so we should also ask whether such modifications alter our result.

It is easy enough to address the latter question. Indeed, if a function $q(\rho)$ satisfies a slow-growth condition near $|\rho|=1$, so does the function $q(\rho)g(\rho)$ and the partial sums of conformal block expansion~\eqref{eq:1dcbexpansion} multiplied by $q(\rho)$. So we can state the straightforward corollary to theorem~\ref{thm:vlad}:
\begin{corollary}
	\label{cor:factors}
	If function $q(\rho)$ is holomorphic in the branched unit $\rho$-disc and satisfies a slow-growth condition near $\tau\in \R$ (recall $\rho=e^{i\tau}$), then we have
	\be
		\label{eq:1dnormlizedcbexpansion}
		\mathrm{bv}\, (q\cdot g) = \sum_{\De} p_\De \mathrm{bv}\, (q\cdot g_\De)\quad\text{in }\cS'(\R). 
	\ee
\end{corollary}
In the example~\eqref{eq:1dcrossing} we have $q(\rho)=z^{-2\De_\f}$ and it satisfies the assumptions of this theorem as can be seen from the identity $z=\frac{4\rho}{(1+\rho)^2}$.

In order to address the questions related to restricting the domain $\cC$ to smaller domains such as $\cC^{st}$, we can prove the following theorem (see Fig.~\ref{fig:DS}).
	\def\SS{S}
\begin{theorem}
	\label{thm:restrictedboundary}
	Let $\mathbb{D}$ be the open unit disk parametrized by $w$ and let $\varphi:w\mapsto\varphi(w)$ be a holomorphic map which maps $\mathbb{D}$ one-to-one onto a domain $ \SS$ inside the cut unit disk of the $\rho$ variable, $ \SS \subset \mathbb{D}\setminus(-1,0]$. Replacing $\rho=\varphi(w)$ in the conformal block expansion~\eqref{eq:1dcbexpansion}, we pull it back to $w\in \mathbb{D}$. Then this pulled-back conformal block expansion in $w$ variable converges on the boundary $|w|=1$ in the sense of distributions (i.e.~when integrated against an arbitrary smooth function on the circle). Furthermore, the same conclusion holds for~\eqref{eq:1dnormlizedcbexpansion} with $q(\rho)=z^{-2\De_\f}$.
\end{theorem}

	\begin{figure}[h!]
	\centering
	\includegraphics[scale=0.8]{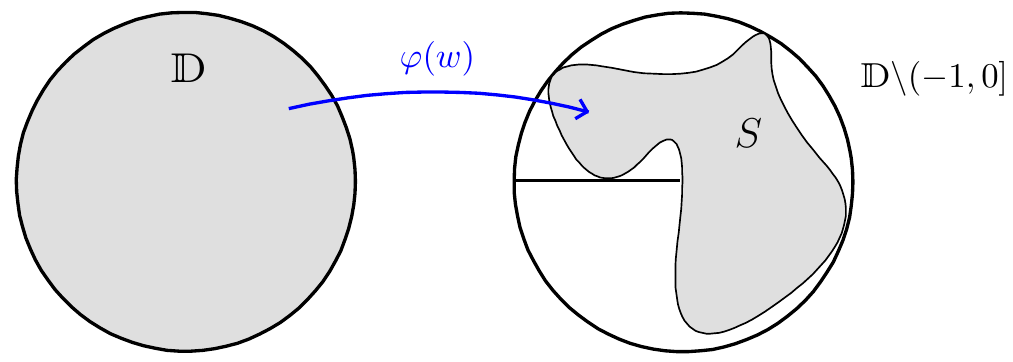}
	\caption{\label{fig:DS} The setting of theorem~\ref{thm:restrictedboundary}. We give one particular example of a possible region $\SS$. In practical applications discussed below $ \SS$ will be either all of $\mathbb{D}\setminus(-1,0]$ or an upper or lower half.}
\end{figure}

	The proof will be based on a simple
	\begin{lemma}
		\label{lemma:functions}
For any one-to-one holomorphic function $\varphi$ from $\mathbb{D}$ onto $ \SS\subset \mathbb{D}\setminus(-1,0]$ there are lower bounds
	\be
	\label{eq:Lind-bound}
	1-|\varphi(w)|&\ge C(1-|w|),\nn\\
	|\varphi(w)|&\ge C'(1-|w|)^2,
	\ee 
	with some $C,C'>0$, and for any $w\in \mathbb{D}$. In other words, the first bound says that $|\varphi(w)|$ cannot approach 1 near the boundary faster than linearly in $w$. Similarly, $|\varphi(w)|$ cannot approach $0$ near the boundary faster than quadratically in $w$. 
	\end{lemma}

	To see why this is intuitively reasonable, consider some model situations. For the first bound, suppose that $\varphi(w)$ has the leading behavior $\varphi_0+ const.(w-w_0)^\alpha$, $|\varphi_0|=1$, near some boundary point $|w_0|=1$. This asymptotics is consistent with~\eqref{eq:Lind-bound} as long as $\alpha\le 1$. The latter condition is implied by the assumption that $\varphi:\mathbb{D}\to \mathbb{D}$: the argument of $w-w_0$ is multiplied by $\alpha$, and for $\alpha>1$ some points will end up outside of the unit circle. A similar check works also for the second bound.

	It should be noted that in practical applications the domain $ \SS$ will typically be either the whole of $\mathbb{D}\setminus(-1,0]$ or its upper or lower half. In these cases the functions $\varphi(w)$ will be explicitly known, and bounds~\eqref{eq:Lind-bound} can be verified by an explicit computation. For completeness, a rigorous general proof of this lemma is given in appendix~\ref{app:lemma}.

	By the first inequality of the lemma, we have the bound $(1-|\rho|)^{-4\De_\f}\leq C'(1-|w|)^{-4\De_\f}$ for some $C'>0$. So the conformal block expansion pulled back to the unit disk $w\in \mathbb{D}$ satisfies the same bounds throughout the disk as the $\rho$-expansion bounds~\eqref{eq:b1}-\eqref{eq:b3}. Recall in particular that $g(\rho)$ is bounded near $\rho=0$ so whatever happens if the boundary of $ \SS$ touches $\rho=0$, as in figure~\ref{fig:DS}, is not important for this part of the argument. Therefore, the first claim of the theorem follows by the same arguments as in section~\ref{sec:distributionalconvergence1d}. There is even one simplification: since the circle is compact, temperedness of distributions having to do with behavior of infinity is of no importance in the case at hand. The space of test functions are $C^\infty$ functions on the unit circle. 
	
	The second claim does not follow immediately because $z^{-2\De_\f}$ blows up near $\rho=0$. However, thanks to the second bound in~\eqref{eq:Lind-bound}, this does not spoil the slow-growth conditions near $|w|=1$. This finishes the proof of the theorem.

Note that we can replace the unit $\rho$-disk by unit $\rho^{1/n}$-disk for some $n$ if we wish to allow the domain parametrized by $w$ to go under the cut. Similarly, the same result can be proven for a wider class of functions $q(\rho)$ than just $z^{-2\De_\f}$. We won't need these generalizations in this paper.

	\begin{figure}[h!]
	\centering
	\includegraphics[scale=0.7]{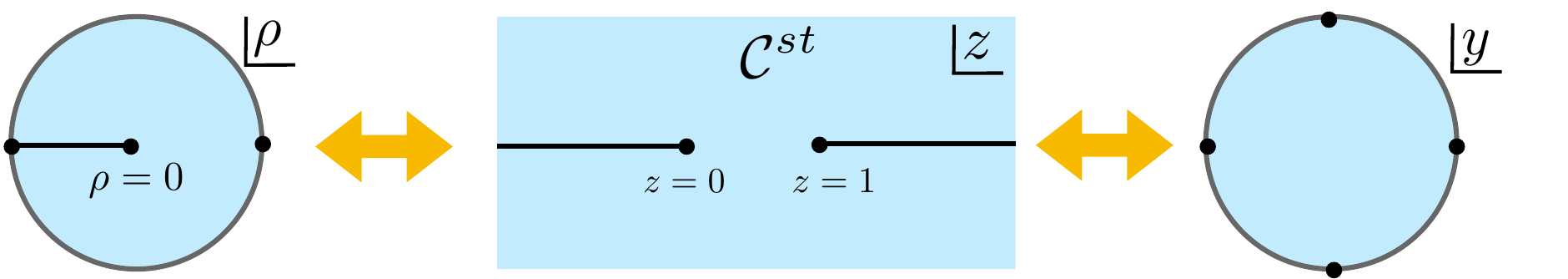}
	\caption{\label{fig:crossing_region} The crossing region $\mathcal{C}^{st}$ and its parametrization using the $\rho$-coordinate and the Zhukovsky $y$-coordinate.}
\end{figure}

\subsection{Analytic functionals}
\label{sec:functionals}
For the first application of theorem~\ref{thm:restrictedboundary}, consider the common region of convergence $\cC^{st}$ of the two OPE channels for the crossing equation given in $z$-coordinate by the cut plane
\be
	\cC^{st}=\C\setminus\p{(-\oo,0]\cup [1,\oo)},
\ee
see figure~\ref{fig:crossing_region}. In $\rho$-variable for either channel it becomes precisely the cut unit disk $\mathbb{D}\setminus(-1,0]$. Following~\cite{Mazac:2016qev}, it is convenient to parametrize domain $\cC^{st}$ via the Zhukovsky map\footnote{The original Zhukovsky (Joukowsky) map $\zeta = y+1/y$ maps the unit circle onto the interval $(-2,2)$. We have $z=1/2+1/\zeta$ so that the unit circle is mapped onto the two cuts $(-\oo,0]\cup [1,\oo)$. The Zhukovsky map is famous in aerodynamics: applying it to offcentric circles one can parametrize airfoil shapes and compute the lift force analytically by conformal invariance of incompressible 2d flows.}
\be
z(y)= \frac{(1+y)^2}{2(1+y^2)}\,.
\ee
This is a holomorphic one-to-one mapping of the unit disk $\mathbb{D}$ onto $\cC^{st}$. Using the function $\varphi(y)=\rho(z(y))$, the $s$-channel conformal block expansion is pulled back to the unit disk of the Zhukovsky variable. Since the region $\cC^{st}$ is symmetric under $z\to 1-z$, the same statement is true for the $t$-channel block expansion (the crossing $z\to 1-z$ corresponds to $y\to -y$). 

We will now apply theorem~\ref{thm:restrictedboundary} with $ \SS=\mathbb{D}\setminus(-1,0]$.
The first conclusion is that the four-point function (both with and without the factor $z^{-2\Delta_\f}$) is a distribution on the boundary of the unit $y$-disk. This statement is only interesting near the points $y=\pm 1$, $y=\pm i$ where the four-point function is singular: on the rest of the boundary it is analytic, as can be shown using the $s$- and $t$-channel expansions.

The second conclusion is that both $s$- and $t$-channel conformal block expansions converge as a distribution on $|y|=1$. This statement is interesting, because in the usual sense each channel converges only on one half of the boundary (the left half for the $s$-channel and the right half for the $t$-channel). 

Distributional convergence has an interesting consequence for the study of the crossing equation using the method of linear functionals~\cite{Rattazzi:2008pe} and in particular for constructing a wide class of functionals satisfying the swapping property of~\cite{Rychkov:2017tpc}. We write the crossing in the usual sum rule form
\be
\label{eq:crossing}
\sum p_\Delta F_\Delta(z)=0,\qquad F_\Delta(z)= z^{-2\De_\f}g_\Delta (z)-(1-z)^{-2\De_\f}g_\Delta(1-z)\,.
\ee
Denote by $F_\Delta(y)$ the same functions pulled back to the unit Zhukovsky disk. They are analytic in the interior and have boundary values $(\text{bv}\,F_\Delta)$ at $|y|=1$. By theorem~\ref{thm:restrictedboundary} we know that~\eqref{eq:crossing} converges on the $|y|=1$ boundary to zero in the sense of distributions. This means that we can integrate it term by term with a smooth function $f(\theta)$:
\be
\label{eq:tbyt}
 \sum p_\Delta \int_0^{2\pi} d\theta (\text{bv}\,F_\Delta)(y=e^{i\theta}) f(\theta)=0\,.
 \ee
The 1d conformal blocks having only logarithmic singularities, the nature of their boundary values is determined by the singularity of prefactors $z^{-2\De_\f}$ and $(1-z)^{-2\De_\f}$. Thus they are ordinary locally integrable functions for $2\Delta_\f<1$, and distributions otherwise. 

Now, let us fix an infinitely smooth $f(\theta)$ on the boundary of the unit $y$-disk. The support of this function may include points in both halfs of the circle, including the points where the four-point function is singular. Consider a linear functional $\a_f$ defined by the formula
\be\label{eq:functional}
	g(y)\mapsto \a_f[g]\equiv\int_0^{2\pi} d\theta (\mathrm{bv}\,g)(y=e^{i\theta}) f(\theta)\,.
\ee
We can write~\eqref{eq:tbyt} equivalently as
\be
\sum p_\De \a_f[F_\De] = 0\,.
\ee
This means, in the terminology of~\cite{Rychkov:2017tpc}, that the functional~\eqref{eq:functional} satisfies swapping property. 

Note that many simple functionals can be rewritten in the form~\eqref{eq:functional}. For example, the derivative evaluation functional $\a_{n,y_0}$
\be
\label{eq:evaluationfunctional}
	g_{n,y_0}(y)\mapsto \a_{n,y_0}[g]\equiv g^{(n)}(y_0)
\ee
for integer $n\geq 0$ and $|y_0|<1$ can be written using Cauchy theorem as\footnote{In Cauchy theorem we integrate over the contour at $|y|=1-\eps$, where the function is analytic. As $\eps\to 0$, the Cauchy kernel tends to $\frac{1}{(e^{i\theta}-y_0)^{n+1}}$ in the $C^\infty$ topology of test functions on the circle, while $g(y)$ tends to $\mathrm{bv}\,g$ in the sense of distributions by theorem \ref{thm:restrictedboundary}. This justifies pushing the contour all the way to the boundary $|y|=1$. }
\be
	\a_{n,y_0}[g]=\frac{n!}{2\pi i}\int_0^{2\pi} d\theta \frac{ie^{i\theta}}{(e^{i\theta}-y_0)^{n+1}}(\mathrm{bv}\,g)(y=e^{i\theta}).
\ee
This coincides with $\a_{f_{n,y_0}}$ with $f_{n,y_0}$ given by
\be\label{eq:evaluationf}
	f_{n,y_0}(\theta)=\frac{n!}{2\pi i}\frac{ie^{i\theta}}{(e^{i\theta}-y_0)^{n+1}}.
\ee

A type of functionals commonly used in analytic functional conformal bootstrap~\cite{Mazac:2016qev,Mazac:2018mdx,Mazac:2018ycv,Kaviraj:2018tfd,Mazac:2018biw,Hartman:2019pcd,Paulos:2019gtx,Mazac:2019shk} can be described as
\be\label{eq:originalcuttouching}
	g(y)\mapsto \a_{h,\G}[g]\equiv \int_{\G} dy h(y)g(y),
\ee
where $h(y)$ is some holomorphic function and $\G$ is a contour in $\mathbb{D}$ which is allowed to have end points on the boundary $|y|=1$. Conditions on $h(y)$ that guarantee the swapping property for $\a_{h,\G}$ were studied in~\cite{Rychkov:2017tpc}. We can try to identify $\a_{h,\G}$ with $\a_{f_{h,\G}}$ where
\be\label{eq:cuttouchingf}
	f_{h,\G}(\theta) = \int_\G dy h(y)f_{0,y}(\theta),
\ee
with $f_{n,y}$ defined in~\eqref{eq:evaluationf}. Unfortunately, if $\G$ ends or starts on $|y|=1$ then for generic $h(y)$ the function $f_{h,\G}(\theta)$ will not be smooth (and so will not be a test function) and thus we have not proven that $\a_{f_{h,\G}}$ is well defined and satisfies the swapping property for this class of functionals. In other words, so far the class of functionals~\eqref{eq:functional} is too small to accommodate the modern results in analytic functional bootstrap. 

However, the swapping conditions of~\cite{Rychkov:2017tpc} require $h(y)$ to decay sufficiently quickly (as some power-law) near the end points of $\G$ that are on $|y|=1$. In this case $f_{h,\G}(\theta)$ is still generically not infinitely smooth, but it will have some finite number of derivatives, i.e.\ we will have $f_{h,\G}(\theta)\in C^k(S^1)$ for some $k>0$. In particular, under the swapping conditions on $h(y)$ derived in~\cite{Rychkov:2017tpc} $k$ is proportional to $\De_\f$. On the other hand, by examining the proof of Vladimirov's theorem~\ref{thm:vlad} given in section~\ref{sec:proof}, we can see that we only use a finite number of semi-norms of the test functions, corresponding to derivatives of order related to the power $M$ in the slow-growth condition~\eqref{eq:slow-growth}, which in turn is related to the dimension $\De_\f$. This implies that in order for the functionals~\eqref{eq:functional} to be well-defined and satisfy the swapping property, we only need $f$ to have $k'$ derivatives with $k'$ proportional to $\De_\f$. 

We thus see that if the functional~\eqref{eq:originalcuttouching} satisfies the swapping conditions derived in~\cite{Rychkov:2017tpc}, then the function~\eqref{eq:cuttouchingf} has $k\propto \De_\f$ derivatives. Similarly, we concluded that our results can be strengthened so that the functional~\eqref{eq:functional} is well defined and satisfies the swapping property if $f$ has $k'\propto \De_\f$ derivatives. This suggests that it is possible to define a space $\cB_{\De_\f}$  of functions on $S^1$ with the following properties. First, we would like $\a_f$ to be well defined and satisfy the swapping property for all $f\in \cB_{\De_\f}$. Furthermore, all functionals used in analytic functional bootstrap should be representable by $\a_f$ with $f\in \cB_{\De_\f}$, i.e.\ we want $f_{h,\G}\in \cB_{\De_\f}$ for all $h$ and $\G$ which satisfy the swapping conditions of~\cite{Rychkov:2017tpc}.

As alluded to above, the first approximation to the space $\cB_{\De_\f}$ is $C^k(S^1)$ with appropriately chosen $k$. However, this seems too coarse, since $k$ is a discrete parameter, while $\De_\f$ is continuous. Moreover, not all the points $y$ with $|y|=1$ are equal---there are special points $y=\pm 1,\pm i$, where the correlator might have a singularity that needs to be controlled, but at all other points we know from crossing that the correlator is smooth (but this does not imply that the conformal block expansion converges there pointwise). It would be interesting to find the appropriate definition for $\cB_{\De_\f}$ since it would provide a uniform description of all functionals suitable for analyzing the crossing equation. We leave these questions for future work.
%
%

\subsection{Dispersion relation in cross-ratio space and the discontinuity}
\label{sec:dispersion}

For a second application, we consider the upper half-plane in $z$ variable. This region is a subset of $\cC^{st}$ and thus we can again use theorem~\ref{thm:restrictedboundary} (this time with $ \SS$ being the upper half of $\mathbb{D}$) to conclude that both $s$- and $t$- conformal block expansions converge as distributions on the boundary of unit disk in the variable $w=\frac{z-i}{z+i}$. 
This boundary minus one point is smoothly mapped to the real line in $z$-plane, and so both $s$- and $t$-channels also converge as distributions on the real line $\R$ in $z$-plane when approached from above.  
By repeating the same arguments for the lower half-plane mapped to the unit disk via $\tl w=\frac{z+i}{z-i}$, we find that both channels converge as distributions on the real line in $z$-plane when approached from below. 

Let us now see how this kind of arguments can be used to write rigorous dispersion relations and give a proper definition of discontinuity (including the point at infinity). Let $z_0$ be a point in the upper half-plane, and $C$ and $\tl C$ be contours in the upper and lower half-planes, with $C$ surrounding $z_0$. Then we have  
\begin{align}
\label{eq:disp0}
g(z_0)&=\frac 1{2\pi i }\oint_C \frac{dz}{z-z_0} g(z)\,,\nn\\
0&=\frac 1{2\pi i }\oint_{\tl C} \frac{dz}{z-z_0} g(z)\,.
\end{align}
Intuitively, to derive the dispersion relation we push $C$ and $\tl C$ to the real axis and infinity, and take the difference of the two equations, which gives a dispersion relation
\begin{equation}
\label{eq:disp}
\quad g(z_0)=\frac 1{2\pi i }\int_{-\infty}^\infty \frac{dx}{x-z_0} {\rm Disc}\, g(x)+\text{contribution at infinity}\,,
\end{equation}
where ${\rm Disc}\, g(x)$ is the difference in two limits of $g(z)$. Contribution at infinity cannot be generally computed in this approach, unless one has some information about the asymptotics of $g(z)$ as $z\to\infty$.

Let us now turn this reasoning into a rigorous dispersion relation, including the contribution at infinity. First of all we pull Eqs. \eqref{eq:disp0} to the unit discs of $w$ and $\tl w$ variable, which gives: 
\begin{align}
\label{eq:disp1}
g(w_0)&=\frac 1{2\pi i } (w_0-1)\oint_C \frac{dw}{w-w_0} \frac{g_+(w)}{w-1},\nn\\
0&=\frac 1{2\pi i } (w_0-1) \oint_{\tl C} \frac{d\tl w/\tl w^2}{\tl w^{-1}-w_0} \frac{g_-(\tl w)}{\tl w^{-1}-1} \,.
\end{align}
where we denoted by $g_+(w)$, $g_-(\tl w)$ the function $g(z)$ from the upper/lower half-plane pulled to the corresponding unit disk. Then we push the contours $C$, $\tl C$ to $|w|=1$, $|\tl w|=1$ and get:
\begin{align}
\label{eq:disp2}
g(w_0)&=\frac 1{2\pi i } (w_0-1)\oint_{|w|=1} \frac{dw}{w-w_0} {\rm bv}\frac{g_+(w)}{w-1},\nn\\
0&=\frac 1{2\pi i } (w_0-1) \oint_{|\tl w|=1}  \frac{d\tl w/\tl w^2}{\tl w^{-1}-w_0} {\rm bv} \frac{g_-(\tl w)}{\tl w^{-1}-1} \,.
\end{align}
Notice that we have to include the singular factors $1/(w-1)$ and $1/(\tl w-1)$, arising due to the transformation of the measure $dz$, under the ``bv'' sign. Since these factors are power-like, the resulting limiting boundary values exist as distributions also in presence of these factors. Finally we take the difference of the two equations and we get:
\begin{align}
\label{eq:disp3}
g(w_0)&=\frac 1{2\pi } (w_0-1)\int_{0}^{2\pi} \frac{d\theta\,e^{i\theta} }{e^{i\theta}-w_0} D(\theta),\\
D(\theta)&=\left.{\rm bv}\frac{g_+(w)}{w-1} \right|_{w=e^{i\theta}} -  \left.{\rm bv} \frac{g_-(\tl w)}{\tl w^{-1}-1}  \right|_{\tl w=e^{-i\theta}}\,.
\end{align}
Here $D(\theta)$ is a distribution on the unit circle, which plays the role of a rigorously defined discontinuity, including the point $z=\infty$ mapped to $\theta=0$. For points away from $\theta=0$ and $\theta=2\pi$ we can pull the factors $1/(w-1)$ and $1/(\tl w-1)$ from under $\mathrm{bv}$ and $D(\theta)$ becomes just
\be
	D(\theta)=\frac{1}{e^{i\theta}-1}\mathrm{Disc}\, g(x=-\cot \tfrac{\theta}{2}), \qquad \theta\neq 0,2\pi.
\ee
Here $\mathrm{Disc}\, g(x=-\cot \tfrac{\theta}{2})$ is defined as $\left.{\rm bv}{g_+(w)} \right|_{w=e^{i\theta}} -  \left.{\rm bv} {g_-(\tl w)} \right|_{\tl w=e^{-i\theta}}$, which is equivalent to taking the boundary values in $z$-space from above and below the real axis, which is simply the intuitive definition of discontinuity. Using this value of $D(\theta)$ in~\eqref{eq:disp3} and changing back to $x$ variable, we recover~\eqref{eq:disp}. So we see that~\eqref{eq:disp3} is indeed an analogue of~\eqref{eq:disp}. However, using $D(\theta)$ allows us to rigorously include the contribution at $x=\oo$.

An intuitive way to think about this construction is that it defines $\mathrm{Disc}\,g(x)$ as a distribution on a class of test functions $\cS_0(\R)$ larger than $\cS(\R)$. The space $\cS_0(\R)$ consists of smooth functions $f(x)$ such that $f(1/x')$ is smooth and vanishing at $x'=0$. Pairing with $\mathrm{Disc}\,g(x)$ is defined by the formula
\be
	\int dx f(x)\mathrm{Disc}\,g(x)\equiv-2\int d\theta e^{i\theta} \tl f(\theta) D(\theta),
\ee
where 
\be\label{eq:iso}
	\tl f(\theta) \equiv \frac{1}{e^{i\theta}-1}f(-\cot \tfrac{\theta}{2})
\ee
is a smooth function on the circle parametrized by $\theta$.\footnote{This equation established isomorphism between $\cS_0(\R)$ and $C^\oo(S^1)$ in the sense that $f\in \cS_0(\R)$ if and only if $\tilde f\in C^\oo(S^1)$.} With this definition we can write the dispersion relation~\eqref{eq:disp3} as
\be\label{eq:disp4}
	g(z_0)=\frac 1{2\pi i }\int_{-\infty}^\infty \frac{dx}{x-z_0} {\rm Disc}\, g(x),
\ee
since the Cauchy kernel $\frac{1}{x-z_0}$ belongs to our new class of test functions. In this language our results imply that both $\mathrm{Disc}$ of the four-point function and $\mathrm{Disc}$ of partial sums of the conformal block expansion are distributions in $\cS'_0(\R)$, and the partial sums converge to the four-point function in this space (i.e.\ discontinuity can be computed term-by-term). 

Let us consider an example. First take $g(z)=\log z$. This is not a good four-point function since it does not satisfy crossing, but it will allow us to clarify the notion of the discontinuity as a distribution and how it can be concretely computed. Going to the $\rho$ variable we easily see that the slow-growth condition is satisfied. For finite $x<0$ we have $\mathrm{Disc}\,g(x)=2\pi i$. This is a distribution in $\cS'(\R)$, but not obviously in $\cS'_0(\R)$. To extend it to $\cS'_0(\R)$ let us write
\be
	\log z = -\lim_{\a\to +0} \ptl_\a z^{-\a}.
\ee
The point here is that $z^{-\a}$ also satisfies a power-law bound and for $\a>0$ the discontinuity
\be
	\mathrm{Disc} \,x^{-\a}=-2 i\sin \pi\a |x|^{-\a}
\ee
is in $\cS'_0(\R)$. We can then obtain $\mathrm{Disc}\,g(x)$ by taking derivative and limit $\a\to +0$.\footnote{Justification for this comes from the limit part of the statement of theorem~\ref{thm:vlad} and (for the derivative) from arguments as in appendix~\ref{app:proofVlad2}.} Pairing $\mathrm{Disc}\,g(x)$ with functions that vanish as $1/x^2$ or faster we get integrals that converge in the usual sense. So we only need to use the limiting construction to define the pairing with $1/x$. We have:
\be
	\int_{-\oo}^{-1}dx \frac{1}{x}\mathrm{Disc}\,g(x) &=-\lim_{\a\to +0} \ptl_\a\int_{-\oo}^{-1}dx\frac{1}{x}\p{-2 i\sin \pi\a |x|^{-\a}}\nn\\
	&=-2 i\lim_{\a\to +0} \ptl_\a\frac{\sin \pi\a}{\a}=0\,,
\ee
where the choice of the integral's upper limit $-1$ is just for convenience since it leads to a simple answer (zero). We can therefore define the distribution $\mathrm{Disc}\,g(x)$ by
\be
	\int dx f(x) \mathrm{Disc}\,g(x) = \int_{-\oo}^{-1} dx (f(x)-f_1x^{-1})2\pi i+\int_{-1}^0 dx f(x)2\pi i.
\ee
where $f_1\equiv\lim\limits_{x\rightarrow\infty}xf(x)$. The dispersion relation~\eqref{eq:disp4} then becomes
\be
	\log z_0 = \int_{-\oo}^{-1} dx \p{\frac{1}{x-z_0}-\frac{1}{x}}+\int_{-1}^0 dx \frac{1}{x-z_0}\,.
\ee
This is easy to verify.

Another example, which we will find useful in section~\ref{sec:Bissi}, is $\mathrm{Disc}\,1$. Naively, this discontinuity must be zero. This is indeed correct, except at $x=\oo$. Indeed, analogously to the above, we have
\be
	1=\lim_{\a\to+0}z^{-\a},
\ee
so
\be
\int_{-\oo}^{-1}dx \frac{1}{x}(\mathrm{Disc}\,1)(x) &=\lim_{\a\to +0} \int_{-\oo}^{-1}dx\frac{1}{x}\p{-2 i\sin \pi\a |x|^{-\a}}\nn\\
&=2 i\lim_{\a\to +0} \frac{\sin \pi\a}{\a}=2\pi i\,,
\ee
and thus 
\be
	\int dx f(x) (\mathrm{Disc}\,1)(x) = 2\pi i f_1,
\ee
where as before $f_1=\lim_{x\to \oo} xf(x)$.

\section{Scalar four-point functions in higher dimensions}
\label{sec:higherd}

We will now generalize our results to general scalar four-point functions in any number of dimensions $d$.  This generalization is mostly technical, and all the conceptual points were already explained in section~\ref{sec:1dcase}. Our strategy is therefore very similar: first we will introduce analogues of the expansions~\eqref{eq:1dcbexpansion} and~\eqref{eq:1dscalingexpansion}, and then use these expansions to prove bounds on the correlation function and partial sums of the conformal block expansion. Finally, we will apply a higher-dimensional version of Vladimirov's theorem~\ref{thm:vlad} to conclude that the conformal block expansion converges in the sense of distributions on the boundary of the region $|\rho|,|\bar\rho|<1$.

\subsection{Conformal block expansion}
\label{sec:generalcb}
We consider a correlation function of four not necessarily identical scalar operators $\f_i$ with scaling dimensions $\De_i$,
\be
\label{eq:general4ptfunction}
	\<\f_1(x_1)\f_2(x_2)\f_3(x_3)\f_4(x_4)\> =  
	\frac{1}
	{
		(x^2_{12})^{\frac{\De_1+\De_2}2}
		(x_{34}^2)^{\frac{\De_3+\De_4}2}
	}
		\p{\frac{x^2_{24}}{x^2_{14}}}^{
		\frac{\De_1-\De_2}2}\p{\frac{x^2_{14}}{x^2_{13}}}^{\frac{\De_3-\De_4}2}g_{1234}(\rho,\bar \rho),
\ee
which is a simple generalization of~\eqref{eq:4pt}. The subscript $1234$ on $g_{1234}$ indicates that it relates to the four-point function of $\phi_1,\ldots,\phi_4$. The function $g_{1234}(\rho,\bar\rho)$ has a conformal block expansion of the form
\be\label{eq:generalscalarcbexpansion}
	g_{1234}(\rho,\bar\rho)=\sum_{\cO} \lambda_{12\bar\cO}\lambda_{43\cO} g_{\De,J}(\rho,\bar\rho),
\ee
where we sum over primaries $\cO$ in $\f_1\times\f_2$ OPE, $\lambda$'s are the three-point coefficients, $\De, J$ are the spin and dimension of $\cO$, and $g_{\De,J}(\rho,\bar\rho)$ are the conformal blocks. The conformal blocks also depend implicitly on $\De_{12}=\De_1-\De_2$ and $\De_{34}=\De_3-\De_4$.

We would like to show that the function $g_{1234}(\rho,\bar\rho)$ satisfies a powerlaw bound as $\rho$ and $\bar\rho$ approach the boundaries of their respective unit disks. We would also like to show that partial sums of the conformal block expansion~\eqref{eq:generalscalarcbexpansion} satisfy a uniform powerlaw bound. We will prove this by relating $g_{1234}(\rho,\bar\rho)$ to the four-point function where operators are inserted symmetrically with respect to the origin~\cite{Pappadopulo:2012jk}.

Let us focus on configurations when all points $x_i$ lie in the 2-plane $P$ defined by $x^\mu=0$ for $\mu>2$. It is convenient to introduce complex coordinates $y,\bar y$ in this plane
\be
	y=x^1+ix^2,\quad \bar y=x^1-ix^2.
\ee
Notice that in Euclidean configurations (i.e.\ when $x^\mu$ are real) we have $\bar y=y^*$. Using the notation $\f_i(y,\bar y)$ for operator insertions in $P$ parametrized by $y,\bar y$, we consider for $\bar\rho=\rho^*$
a symmetrically-inserted four-point function
\be\label{eq:rhocf}
	\tl g_{1234}(\rho,\bar \rho) = (\rho\bar\rho)^{\frac{\De_1+\De_2}{2}} \<\f_1(-\rho,-\bar\rho)\f_2(\rho,\bar\rho)\f_3(1,1)\f_4(-1,-1)\>\,,
\ee
where the factor $ (\rho\bar\rho)^{\frac{\De_1+\De_2}{2}}$ is inserted for further convenience (basically to make Eq.~\eqref{eq:higherdscalingexpansion} look maximally nice)\,.
For operators inserted as shown, the meaning of $\rho$ in~\eqref{eq:rhocf} and~\eqref{eq:general4ptfunction} is the same, justifying the notation. Evaluating also the prefactor in~\eqref{eq:general4ptfunction}, we find the following relation between $\tl g_{1234}$ and $g_{1234}$:\footnote{Both $\tl g_{1234}$ and $g_{1234}$ depend on $\rho,\bar\rho$ and both can pretend to be called the conformally invariant part of the general four-point function. One could switch from one convention to the other by changing the prefactor in~\eqref{eq:general4ptfunction}. We will still express our final results in terms of $g_{1234}$, since Eq.~\eqref{eq:general4ptfunction} is the most standard convention.}
\be
	\label{eq:gtlgrelation}
	\tl g_{1234}(\rho,\bar\rho)={2^{-\De_1-\De_2-\De_3-\De_4}}\p{\frac{(1+\rho)(1+\bar\rho)}{(1-\rho)(1-\bar\rho)}}^{\half(\De_{12}-\De_{34})} g_{1234}(\rho,\bar\rho).
\ee
For $\bar\rho=\rho^*$~\eqref{eq:rhocf} is a Euclidean configuration, radial quantization of which~\cite{Pappadopulo:2012jk,Fitzpatrick:2012yx,Hogervorst:2013sma,Hartman:2015lfa} gives the following absolutely convergent expansion for $|\rho|=|\bar\rho|<1$
\be
	\label{eq:higherdscalingexpansion}
	\tl g_{1234}(\rho,\bar\rho) = \sum_{\psi} \tl \lambda_{12\psi}\tl\lambda_{43\bar\psi} \rho^{h}{\bar\rho}^{\bar h},
\ee
where we sum over eigenstates $\psi$ of dilatations and planar rotations in radial quantization, and $h,\bar h$ are appropriate combinations of the corresponding eigenvalues. Since it converges absolutely for $|\rho|=|\bar\rho|<1$ when $\bar\rho=\rho^*$,
it also does so for independent $\rho$ and $\bar\rho$ when $|\rho|,|\bar\rho|<1$. Furthermore, the conformal block expansion~\eqref{eq:generalscalarcbexpansion} can be understood as a reorganization of expansion~\eqref{eq:higherdscalingexpansion} by
grouping $\psi$ into conformal families.

\subsection{Bounds on $g(\rho,\bar\rho)$ and partial sums of the conformal block expansion}
\label{sec:generalbounds}
\def\mm{\hspace{0.05em}}
Consider the following analogues of~\eqref{eq:rhocf},\eqref{eq:higherdscalingexpansion} where two pairs of operators are hermitean conjugates of each other:
\be
\label{eq:an1}
	\tl g_{12\bar 2\mm\bar 1}(\rho,\bar\rho) &
	=  \sum_{\psi} \tl \lambda_{12\psi}\tl\lambda_{\bar 1\mm\bar 2\mm\bar\psi} \rho^{h}{\bar\rho}^{\bar h}
	=\sum_{\psi} |\tl \lambda_{12\psi}|^2\rho^{h}{\bar\rho}^{\bar h},\\
	\label{eq:an2}
	\tl g_{\bar 4\mm\bar 334}(\rho,\bar\rho) &=  \sum_{\psi} \tl \lambda_{\bar 4\mm\bar 3\psi}\tl\lambda_{43\bar\psi} \rho^{h}{\bar\rho}^{\bar h}
	=  \sum_{\psi} |\tl\lambda_{43\bar\psi}|^2 \rho^{h}{\bar\rho}^{\bar h},
\ee
where we use $\bar 1$, etc., to denote three-point coefficients of hermitian conjugates $\phi_1^\dagger$, etc.. As shown, because of $\tl\lambda_{\bar 1\mm\bar 2\bar\psi}=(\tl\lambda_{12\psi})^*$ and $\tl\lambda_{\bar 4\bar 3\psi}=(\tl\lambda_{43\bar\psi})^*$, these two expansions have non-negative real coefficients. Furthermore, estimating by absolute value and applying Cauchy-Schwarz, we can bound~\eqref{eq:higherdscalingexpansion} in terms of~\eqref{eq:an1},~\eqref{eq:an2}:
\be
\label{eq:scalarbound1}
	|\tl g_{1234}(\rho,\bar \rho)| \leq 
\sum_{\psi} |\tl \lambda_{12\psi}||\tl\lambda_{43\bar\psi}| r^{h+\bar h}\le \bigl [\tl g_{12\bar 2\bar 1}(r,r)\tl g_{\bar 4\bar 334}(r,r)\bigr]^{1/2}
\ee
where $r=\max(|\rho|,|\bar\rho|)$.\footnote{We also have a more nuanced bound by $\bigl[\tl g_{12\bar 2\bar 1}(|\rho|,|\bar\rho|)\tl g_{\bar 4\bar 334}(|\rho|,|\bar\rho|)\bigr]^{1/2}$ but we won't need it.} Note that the same bound holds if we replace the sum over $\psi$ by a sum over a subset of all allowed $\psi$'s. This, similarly to the argument in section~\ref{sec:distributionalconvergence1d}, implies that the partial sums of expansions~\eqref{eq:generalscalarcbexpansion} and~\eqref{eq:higherdscalingexpansion} satisfy the same bound~\eqref{eq:scalarbound1} (with $\tl g_{1234}$ related to $g_{1234}$ via~\eqref{eq:gtlgrelation} where needed).

To proceed we need a bound on $\tl g_{12\bar 2\bar 1}(r,r)$ and $\tl g_{\bar 4\bar 334}(r,r)$. 
This bound is easy to obtain from the corresponding definition~\eqref{eq:rhocf}. In the limit $r\to1$ two pairs of hermitean conjugate operators approach each other. Using OPE between the approaching pairs, we get a leading asymptotics for the correlator.\footnote{This can be equivalently formulated via crossing symmetry in $z$ space and then transforming to the $\rho$ space, as in section~\ref{sec:distributionalconvergence1d}.} This implies a bound of the same functional form as the leading asymptotics times a constant. The resulting bounds have the form:
\be
	\tl g_{12\bar2\bar 1}(r, r)&\leq C(1-r)^{-2\De_1-2\De_2},\\
	\tl g_{\bar 4\bar 334}(r, r)&\leq C(1-r)^{-2\De_3-2\De_4},
\ee
with some $C>0$. Notice that there is no blowup as $r\to 0$ since it's overcome by the prefactor in~\eqref{eq:rhocf}. Combining these with~\eqref{eq:scalarbound1} we find
\be
	|\tl g_{1234}(\rho,\bar \rho)|\leq C (1-r)^{-\De_1-\De_2-\De_3-\De_4},
\ee
and finally via~\eqref{eq:gtlgrelation} we get a bound for $g_{1234}$
\be
	\label{eq:generalpowerbound}
	|g_{1234}(\rho,\bar\rho)|\leq C'(1-r)^{-\De_1-\De_2-\De_3-\De_4-|\De_{12}-\De_{34}|},\qquad r=\max(|\rho|,|\bar\rho|),
\ee
for some $C'>0$. Again, the same bound with the same $C'$ holds for the partial sums of expansions~\eqref{eq:generalscalarcbexpansion} and~\eqref{eq:higherdscalingexpansion}.

We repeat the logic of this argument. The key idea is to use OPE in the cross channel to infer the leading singularity of the correlator and then to argue that a similar bound holds throughout the range $|\rho|,|\bar \rho|<1$. This does not work directly for $g_{1234}$, but only for 4pt functions with non-negative $\rho$,$\bar\rho$ expansion coefficients, such as $g_{12\bar 2\bar 1}$ and $g_{\bar 4\bar 3 34}$. So we run the argument for those, and recover the general case by Cauchy-Schwarz.

\subsection{Vladimirov's theorem}

Now that we have the bound~\eqref{eq:generalpowerbound} we would like to use a higher-dimensional version of Vladimirov's theorem~\ref{thm:vlad} to argue for the distributional convergence of conformal block expansion~\eqref{eq:generalscalarcbexpansion}.

\begin{theorem}
	\label{thm:vlad2}
	Consider $\C^{N}=\C^n\times \C^d$ with coordinates $w_k$ on $\C^n$ and $u_k=x_k+i y_k$ on $\C^d$. Let $U$ be an open subset of $\C^n$ and let $M=U\times \R^d$ be the manifold defined by $w\in U, \, y_k=0,\, k=1\ldots d$. Let $V$ be a convex open cone in $\R^d$ with vertex at $y=0$ that doesn't contain $y=0$.  Let $\cW$ be the subset of $\C^{N}$ for which $y\in V$, $|y_k|<a$ for some $a>0$, and $w\in U$. Let $g(w,u)$ be a function holomorphic in $\cW$ that satisfies the slow-growth condition near $M$\footnote{More precisely, we'd like to have this condition satisfied uniformly on compact subsets $w\in \cK\subset U$ with $C,L,K$ allowed to depend on $\cK$.\label{ft:compactfootnote}}
	\be
		\label{eq:slow-growth2}
		|g(w,u)|\leq C\p{1+\sum_k x_k^2}^L \p{\sum_k y_k^2}^{-K}.
	\ee
	Finally, let $v$ be a vector in $V$. Then for fixed $w$ the boundary value
	\be
		(\mathrm{bv}\,g)(w,x) = \lim_{\e\to+0} g(w,x+iv\e)
	\ee
	exists in $\cS'(\R^d)$ and is independent of the choice of $v$. Furthermore, this boundary value depends on $w$ holomorphically, which means that for any $f\in \cS(\R^d)$ the function $h(w)$ defined by\footnote{Here the integral of course just means the pairing of the distribution with the test function.}
	\be
		h(w)\equiv \int d^dx\, f(x) (\mathrm{bv}\,g)(w,x)
	\ee
	is holomorphic for $w\in U$. Furthermore, suppose that sequence of functions $g_n$ holomorphic in $\cW$ converges to $g$ in $\cW$ pointwise and satisfies the slow-growth condition near $M$ uniformly in $n$. Then for all $w\in U$
	\be\label{eq:generalfnconvergence}
		(\mathrm{bv}\,g_n)(w,x) \to (\mathrm{bv}\,g)(w,x)\quad\text{in }\cS'(\R^d).
	\ee
\end{theorem}
The proof of this theorem is very similar to the proof of theorem~\ref{thm:vlad} given in section~\ref{sec:proof}, and we summarize it in appendix~\ref{app:proofVlad2}. For more general results in this direction see, for example,~\cite{Vladimirov} and~\cite{RealSubmanifolds}.

Let us now apply theorem~\ref{thm:vlad2} to the conformal block expansion~\eqref{eq:generalscalarcbexpansion}. As a first step, we introduce the coordinates $\tau$ and $\bar\tau$ via
\be
	\rho=e^{i\tau},\quad\bar\rho=e^{i\bar\tau}.
\ee
Note that in Euclidean configurations we have $\bar\tau=-\tau^*$. The function $g_{1234}(\tau,\bar\tau)$ as well as the partial sums of~\eqref{eq:generalscalarcbexpansion} are holomorphic functions in the region
\be
	\cW_0=\{(\tau,\bar\tau)|\mathrm{Im}\,\tau,\mathrm{Im}\,\bar\tau>0\},
\ee
which is the universal cover of the product of open unit discs of $\rho$ and $\bar\rho$. Furthermore, the expansion~\eqref{eq:generalscalarcbexpansion} converges absolutely in $\cW_0$. We can apply theorem~\ref{thm:vlad2} in two essentially different ways.

Firstly, we can take $\mathrm{Im}\,\bar\tau$ to zero while keeping $\tau$ fixed. This corresponds to $n=d=1$ case of theorem~\ref{thm:vlad2}, in which $\C^n$ is parametrized by $\tau$ and $\C^d$ by $\bar\tau$. The open set $U$ is then given by $\mathrm{Im}\,\tau>0$ and the cone $V$ is given by $y_1=\mathrm{Im}\,\bar\tau>0$. The set $\cW$ is then
\be\label{eq:simplecW}
	\cW=\{(\tau,\bar\tau)|\mathrm{Im}\,\tau>0,a>\mathrm{Im}\,\bar\tau>0\},
\ee
for some $a>0$, say $a=1$.
The slow-growth condition for $g_{1234}(\tau,\bar\tau)$ and the partial sums of~\eqref{eq:generalscalarcbexpansion} follows from~\eqref{eq:generalpowerbound}. In this way, for each $\tau$ we get a distribution
\be
	(\mathrm{bv}\,g_{1234})(\tau,\mathrm{Re}\,\bar\tau)=\sum_{\cO} \lambda_{12\bar\cO}\lambda_{43\cO} (\mathrm{bv}\,g_{\De,J})(\tau,\mathrm{Re}\,\bar\tau)\quad\text{in }\cS'(\R)
\ee
that is holomorphic in $\tau$. Similarly, we can send $\mathrm{Im}\,\tau$ to $0$ while keeping $\tau$ fixed to get
\be
	(\mathrm{bv}\,g_{1234})(\mathrm{Re}\,\tau,\bar\tau)=\sum_{\cO} \lambda_{12\bar\cO}\lambda_{43\cO} (\mathrm{bv}\,g_{\De,J})(\mathrm{Re}\,\tau,\bar\tau)\quad\text{in }\cS'(\R),
\ee
holomorphic in $\bar\tau$.

\begin{figure}[h!]
	\centering
	\includegraphics[scale=0.5]{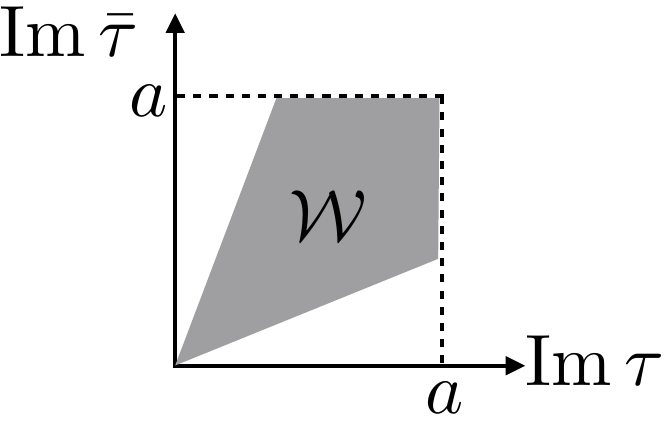}
	\caption{\label{fig:W} The region $\cW$ relevant for the second application of Vladimirov's theorem.}
\end{figure}

Secondly, we can take the simultaneous limit $\mathrm{Im}\,\tau,\mathrm{Im}\,\bar\tau\to 0$. This corresponds to $n=0$ and $d=2$ in theorem~\ref{thm:vlad2}. A small subtlety is that with $\cW$ as in~\eqref{eq:simplecW} the slow-growth condition doesn't follow from~\eqref{eq:generalpowerbound}, since $r$ in~\eqref{eq:generalpowerbound} can approach $1$ even if only one of $\mathrm{Im}\,\tau,\mathrm{Im}\,\bar\tau$ is small. To fix this, choose any $\a<1$ and define
\be
	V=\{(\mathrm{Im}\,\tau,\mathrm{Im}\,\bar\tau)\,|\, \mathrm{Im}\,\tau,\mathrm{Im}\,\bar\tau>0,\, \a^{-1}<\mathrm{Im}\,\tau/\mathrm{Im}\,\bar\tau<\a\}.
\ee
The corresponding $\cW$ has form as in figure~\ref{fig:W}. In this new $\cW$ we have $1-r>C[(\mathrm{Im}\,\tau)^2+(\mathrm{Im}\,\bar\tau)^2]^{1/2}$ for some $C>0$ and the slow-growth condition follows from~\eqref{eq:generalpowerbound}. We therefore conclude the existence of the boundary values and the distributional convergence of the boundary value series:
\be
	(\mathrm{bv}\,g_{1234})(\mathrm{Re}\,\tau,\mathrm{Re}\,\bar\tau)=\sum_{\cO} \lambda_{12\bar\cO}\lambda_{43\cO} (\mathrm{bv}\,g_{\De,J})(\mathrm{Re}\,\tau,\mathrm{Re}\,\bar\tau)\quad\text{in }\cS'(\R^2).
\ee

\subsection{Analytic functionals}

Similarly to the one-dimensional case, we can consider various generalizations. In particular, we have the obvious generalizations of corollary~\ref{cor:factors} and theorem~\ref{thm:restrictedboundary}.

\begin{corollary}
	\label{cor:factors2}
	If function $q(\rho,\bar\rho)$ is holomorphic in the branched unit $\rho,\bar\rho$-polydisc and satisfies the appropriate slow-growth conditions near $\tau,\bar \tau\in \R$ (recall $\rho=e^{i\tau},\bar\rho=e^{i\bar\tau}$), then we have
	\be
	(\mathrm{bv}\,q\cdot g_{1234})(\tau,\mathrm{Re}\,\bar\tau)&=\sum_{\cO} \lambda_{12\bar\cO}\lambda_{43\cO} (\mathrm{bv}\,q\cdot g_{\De,J})(\tau,\mathrm{Re}\,\bar\tau)\quad\text{in }\cS'(\R),\\
	(\mathrm{bv}\,q\cdot g_{1234})(\mathrm{Re}\,\tau,\bar\tau)&=\sum_{\cO} \lambda_{12\bar\cO}\lambda_{43\cO} (\mathrm{bv}\,q\cdot g_{\De,J})(\mathrm{Re}\,\tau,\bar\tau)\quad\text{in }\cS'(\R),\\
	(\mathrm{bv}\,q\cdot g_{1234})(\mathrm{Re}\,\tau,\mathrm{Re}\,\bar\tau)&=\sum_{\cO} \lambda_{12\bar\cO}\lambda_{43\cO} (\mathrm{bv}\,q\cdot g_{\De,J})(\mathrm{Re}\,\tau,\mathrm{Re}\,\bar\tau)\quad\text{in }\cS'(\R^2).
	\ee
\end{corollary}

\begin{theorem}
	\label{thm:restrictedboundary2}
	Let $\mathbb{D}$ be the open unit disk parametrized by $w$ and let $\varphi:w\mapsto\varphi(w)$ be a holomorphic map which maps $\mathbb{D}$ one-to-one onto a domain $ \SS$ inside the cut unit disk of the $\rho$ variable, $ \SS \subset \mathbb{D}\setminus(-1,0]$. Let $\bar\phi$ be a map of the same kind with $ \SS$ replaced by $\bar{ \SS}\subset \mathbb{D}\setminus(-1,0]$. Replacing $\rho=\phi(w),\bar\rho=\bar\phi(\bar w)$ in the conformal block expansion~\eqref{eq:generalscalarcbexpansion}, we pull it back to $w,\bar w\in \mathbb{D}\times\mathbb{D}$. Then this pulled-back conformal block expansion in $w,\bar{w}$ variables converges on the boundaries $|w|=1$, $|\bar w|=1$, or $|w|=|\bar w|=1$ in the sense of distributions. Furthermore, the same conclusion holds if expansion~\eqref{eq:generalscalarcbexpansion} is multiplied by  $q(\rho,\bar \rho)=(z\bar z)^{-\frac{\De_1+\De_2}{2}}$.
\end{theorem}

For example, the discussion of analytic bootstrap functionals in section~\ref{sec:functionals} can be extended to the two-variable case as follows. In Zhukovsky variables $y,\bar y$ the crossing domain $\cC^{st}$ is given by $\mathbb{D}\times \mathbb{D}$. The boundary $\ptl(\mathbb{D}\times \mathbb{D})$ is topologically a 3-sphere $S^3$. This $S^3$ is a disjoint union
\be
	S^3 = (\mathbb{D}\times S^1) \sqcup (S^1 \times \mathbb{D}) \sqcup \mathbb{T}^2,
\ee
where the first solid torus $\mathbb{D}\times S^1$ corresponds to $|y|=1$ and $|\bar y|<1$, the second solid torus corresponds to $|\bar y|=1$ and $|y|<1$, while the torus $\mathbb{T}^2=S^1\times S^1$ corresponds to $|y|=|\bar y|=1$. We have shown that the conformal block expansion in either $s$- or $t$- channel converges in the sense of distributions on each of these boundary components. 

Let us focus on the component $\mathbb{T}^2=S^1\times S^1$. Our results imply that the functionals  $\a_f$ of the form
\be
	g(y,\bar y)\mapsto \a_f[g]\equiv \int_0^{2\pi}\int_0^{2\pi} d\theta d\bar\theta f(\theta,\bar\theta) g(y=e^{i\theta},\bar y=e^{i\bar\theta})
\ee
where $f(\theta,\bar\theta)$ is a smooth function, satisfy the swapping property. As in section section~\ref{sec:functionals}, by taking $f$ to be the Cauchy kernel
\be
	f_{m,n;y_0,\bar y_0}(\theta,\bar\theta) = \frac{m!n!}{(2\pi)^2} \frac{e^{i\theta}}{(e^{i\theta}-y_0)^{m+1}}\frac{e^{i\bar\theta}}{(e^{i\bar\theta}-\bar y_0)^{n+1}},
\ee
we can reproduce the evaluation functionals $\a_{m,n;y_0,\bar y_0}$ 
\be
	g(y,\bar y)\mapsto \a_{m,n;y_0,\bar y_0}[g]\equiv \ptl_y^m\ptl_{\bar y}^n g(y_0,\bar y_0).
\ee
We can again ask about the space of functions $f$ for which the functional $\a_f$ satisfies the swapping property and try to see if this space is large enough to incorporate the functionals that are useful in analytic conformal bootstrap. Just as in section~\ref{sec:functionals}, we leave these questions for future work.

\subsection{Spinning operators}

Another natural generalization available in higher dimensions is to operators with spin. In cross-ratio space this question is somewhat non-canonical due to the freedom of choosing the tensor structures for spin indices, which is similar to the freedom of selecting the prefactor in~\eqref{eq:general4ptfunction}. Nevertheless, it is clear that for reasonable choices of the basis of tensor structures, the four-point functions of spinning operators should satisfy similar power-law bounds in cross-ratio space. For example, one could use equation~\eqref{eq:rhocf} with $\phi_i$ replaced by plane-rotation eigencomponents of some spinning operators $\cO_i$, and the arguments of sections~\ref{sec:generalcb} and~\ref{sec:generalbounds} would still go through. This would correspond to using the ``conformal frame'' basis of four-point structures~\cite{Kravchuk:2016qvl}, which is related to all reasonable choices of tensor structures by matrices which themselves satisfy power-law bounds.\footnote{There is a subtlety for $z=\bar z$, in which case the transition matrices to/from conformal frame basis become singular. These singularities are canceled by special conditions satisfied by four-point functions in conformal frame basis near this locus (see appendix A of~\cite{Kravchuk:2016qvl} and appendix D of~\cite{Karateev:2019pvw}).} While it would be a good exercise to explicitly repeat our arguments in the case of spinning correlators, we do not do it in this paper for the sake of space.

\subsection{Single-variable dispersion relation for the four-point function in $d\ge 2$}
\label{sec:Bissi}
This section generalizes section \ref{sec:dispersion} to the case of two cross-ratios $z,\bar{z}$. Ref.~\cite{Bissi:2019kkx} presented a single-variable dispersion relation recovering the four-point function in terms of its discontinuities. We will state their story in our language, clarifying some issues. Consider the four-point function satisfying the crossing equation
\be
	\label{eq:crossing_disp}
F(z,\bar z)=F(1-z,1-\bar z) = (z \bar z)^{-\De_\f} F(1/z,1/\bar{z})\,,
\ee
where $F(z,\bar z)= (z \bar z)^{-\De_\f}g(z,\bar z)$ and the third equation corresponds to the $u$-channel. This channel representation does not exist for a general 1d four-point function considered in section \ref{sec:1dvariants}.

Ref.~\cite{Bissi:2019kkx} considers a dispersion relation for the function $F(z,\bar z)$ using the discontinuity w.r.t.~$z$ and keeping $\bar z$ fixed. In our language this dispersion relation would be written in the form 
\begin{equation}
\label{eq:dispgen}
\quad F(z,\bar z)=\frac 1{2\pi i }\int_{-\infty}^\infty \frac{dx}{z'-z} \mathop{{\rm Disc}}\limits_{z'}\, F(z',\bar z)
\end{equation}
where the discontinuity has to be understood in a distributional sense, including the contribution at infinity, as discussed in section \ref{sec:dispersion}.

Then the question arises how to compute the discontinuity. There are three cases: $-\infty<z'\le 0$, $1\le z'< +\infty$, and $z'=\infty$. In the first case we can use the $s$-channel conformal block decomposition, which converges in the sense of distributions (in fact in ordinary sense for $z'<0$). The discontinuity at $z'\ge 1$ is reduced to the one at $z'\le 0$ via the first crossing equation in \eqref{eq:crossing_disp}.\footnote{It is also possible to compute the discontinuity at $z\ge 1$ by summing the $s$-channel conformal block expansion since by our results it converges on this cut in the sense of distributions. Ref.~\cite{Bissi:2019kkx} mentions this result in footnote 1, attributing it to Mack~\cite{Mack:1976pa}. This is not correct: Mack's paper studies distributional convergence of OPE expansion in position space, not in the cross-ratio space as needed here. In due fairness, footnote 1 is not central for \cite{Bissi:2019kkx}, being only used in section 4.2.2.
	
We note in passing that Mack \cite{Mack:1976pa} relied on validity of Wightman axioms and rather non-trivial representation theory. It is only in~\cite{paper2,paper3} that we will show, for the first time, how some of Mack's assumptions follow from more mundane Euclidean CFT rules. In comparison, our arguments here are very elementary and rely only on the well-established properties of the conformal block expansion.}

One can try to fix the contribution at infinity using the $u$-channel conformal block expansion, which determines the behavior of the correlator at $z'=\infty$. Let us assume that 
\be
\label{eq:asu}
F(z',\bar z) = 1+O((z')^{-\tau/2})\,.
\ee 
Ref.~\cite{Bissi:2019kkx} argued this by appealing to the second crossing relation in \eqref{eq:crossing_disp}, expanding $F(1/z,1/\bar z)$ in conformal blocks, keeping only the unit operator and dropping all the other operators which seem to be naively suppressed by $(1/z)^{\tau/2}$ where $\tau=\min(\Delta-\ell)$ is the minimal twist, assumed positive. This reasoning includes a subtlety, see below. But assuming \eqref{eq:asu} we can argue that, in the language of section~\ref{sec:dispersion},
\be
	\mathop{{\rm Disc}}\limits_{z}\,F(z,\bar z) = (\mathrm{Disc}\, 1)(z) + \mathop{{\rm Disc}'}\limits_{z}\,F(z,\bar z),
\ee
where $\mathrm{Disc}\, 1$ was computed in section~\ref{sec:dispersion} and $\mathop{{\rm Disc}'}\limits_{z}\,F(z,\bar z)$ is a distribution that is represented near $z=\oo$ by an ordinary function. In other words, ${\rm Disc}'$ is the discontinuity ``without the contribution at $\oo$.''\footnote{Note, however, that we can only unambiguously define such discontinuity because of~\eqref{eq:asu}. For example, this is not possible for $\log z$ example from section~\ref{sec:dispersion}.}

Using this decomposition of $\mathop{\mathrm{Disc}}\limits_{z}\,F(z,\bar z)$, one obtains from~\eqref{eq:dispgen} a dispersion relation in the form given by \cite{Bissi:2019kkx}
\begin{equation}
\label{eq:dispgen1}
\quad F(z,\bar z)=1+ \Bigl(\frac 1{2\pi i }\int_{-\infty}^0 \frac{dz'}{z'-z} \mathop{{\rm Disc}'}\limits_{z'}\, F(z',\bar z)+(z,\bar z\to 1-z,1-\bar z)\Bigr)\,,
\end{equation}
where, as mentioned above, the discontinuity $\mathrm{Disc}'$ does not include the contribution at infinity that is instead explicitly included as ``$1+$'', and we used crossing symmetry to account for discontinuity on the cut $[1,+\oo)$.

Note that independently of the assumption~\eqref{eq:asu}, our results imply that $\mathop{\mathrm{Disc}}\limits_{z} F(z,\bar z)$ can be computed term-by term in conformal block expansion (including the contribution at infinity), and then used in~\eqref{eq:dispgen}, although it is not guaranteed that the decomposition~\eqref{eq:dispgen1} exists in that case.

Let us now discuss the subtlety in the asymptotics \eqref{eq:asu}. Upon a closer look, this asymptotic is only justified provided that $z$ and $\bar z$ 
belong to the different halfplanes of the region $\cC^{st}$, i.e.~if ${\rm Im}\,z$ and ${\rm Im}\,\bar z$ have opposite sign. This is because the $u$-channel conformal block expansion stops converging when $z$ crosses the cut $(0,1)$ and moves into the same half-plane as $\bar z$. Thus, if $\bar z$ is fixed, asymptotics \eqref{eq:asu} is rigorously true only on one of the two arcs at infinity $z$. The asymptotics on the second arc is somewhat similar to the Regge limit asymptotics, in the sense that $1/z$ goes through the $s$-channel cut and then is sent to zero (while, unlike in the Regge limit, $\bar z$ stays fixed). 

There are two ways around this difficulty. One way is to take $\bar z\in (0,1)$ real. Then, by our results, the $u$-channel OPE expansion converges in the sense of distributions on both arcs. In this case the asymptotics \eqref{eq:asu} is true provided that the error term is understood in the sense of distributions, and it goes to zero as $z\to \infty$. Since a zero distribution is a zero function, we recover the dispersion relation \eqref{eq:dispgen1}.

The second way around the difficulty is to apply the dispersion relation in perturbation theory around a mean field theory, which was in fact the main focus of \cite{Bissi:2019kkx}. In their case the zeroth order term satisfies the asymptotics \eqref{eq:asu} by inspection, while perturbative corrections have an even better behavior. The use of dispersion relation in such a limited context is justified.

\section{Conclusions}
\label{sec:conclusions}

In this work we studied the properties of the conformal block expansion on the boundary of its region of convergence. We showed that both the correlation functions and conformal blocks can be interpreted as distributions on this boundary, and that the conformal 
block expansion converges in the space of distributions. We have proven these results in one- and higher-dimensional cases for correlators of scalar operators, but the extension to general spinning four-point functions is straightforward.

An important feature of our analysis is that we did not rely on anything but the modern Euclidean bootstrap axioms. Specifically, we essentially only used the reality properties of OPE coefficients and the usual convergence properties of the conformal block expansion. There is a growing consensus that the Euclidean bootstrap axioms provide a good conceptual and practical definition for CFTs. Their conceptual appeal is due to them being rooted in cutting-and-gluing properties of Euclidean path integrals, which is a natural expected consequence of locality. The practical utility of these axioms has been demonstrated by the numerical conformal bootstrap studies, which have yielded extremely precise values of critical exponents and other parameters in various strongly-coupled CFTs such as the 3d Ising CFT and the $O(2)$ model (see \cite{Kos:2016ysd, Chester:2019ifh} for the most precise determination to date). These values are in agreement with a plethora of other completely independent methods (most notably Monte Carlo simulations and the $\eps$-expansion).

Our results are important for understanding the nature of conformal correlation functions in Lorentzian signature. Indeed, as we show in appendix~\ref{app:lorentz}, the best one can guarantee in general configurations in Lorentzian signature is that the conformal cross-ratios are on the boundary of the region of convergence for one of the OPE channels. It is thus important to understand the value of CFT four-point functions on this boundary. We have shown that the conformal block expansion converges there in distributional sense, which gives a practical way for computing correlation functions. For example, we can now imagine collecting numerical OPE data for 3d Ising CFT as in~\cite{Simmons-Duffin:2016wlq} and using it to compute pairings of the boundary value of $\<\s\s\s\s\>$ four-point function with various tests functions.

One important byproduct of our results, which we discuss in section~\ref{sec:functionals}, is a hint at a uniform description of the space of functionals with which we can probe the crossing equation. Starting with numerical conformal bootstrap~\cite{Rattazzi:2008pe}, it has become standard to disprove the existence (under certain spectral assumptions) of solutions to the crossing equation by exhibiting functionals that separate the left-hand side of the crossing equation from the right-hand side. In numerical bootstrap (see~\cite{Poland:2018epd} for review) these functionals are finite combinations of evaluation functionals $\a_{n,y}$~\eqref{eq:evaluationfunctional}, while in more recent analytical functional bootstrap~\cite{Mazac:2016qev,Mazac:2018mdx,Mazac:2018ycv,Kaviraj:2018tfd,Mazac:2018biw,Hartman:2019pcd,Paulos:2019gtx,Mazac:2019shk} the appropriate functionals are given by contour integrals $\a_{h,\G}$~\eqref{eq:originalcuttouching}. Having a uniform description of a sufficiently large class $\cB_{\De_\f}$ of functionals (that in particular would include $\a_{n,y}$ and $\a_{h,\G}$) would allow us to formulate and hopefully answer some interesting conceptual questions. For example, 
\begin{itemize}
	\item is it true that for any spectral assumption for which there is no solution to crossing equation there exists a functional in $\cB_{\De_\f}$ that disproves the existence of a solution? 
	\item Is it true that when the spectral assumption is not ``extremal,'' this functional can be taken as a finite linear combination of evaluation functionals? (In other words, is numerical conformal bootstrap complete?) 
	\item When the spectral assumption is extremal, is it true that there exists a unique extremal functional? 
\end{itemize}

Most practitioners would probably guess that the answer to these three questions should be ``yes'', ``yes'' and ``generically yes''. To put this intuition on firm footing we need first of all understand better the space $\cB_{\De_\f}$ and the appropriate topology on this space. 
Answering these questions will be important for advancing our analytical understanding of conformal bootstrap.

\section*{Acknowledgments}

Some of our results were first presented in a series of four lectures at the CEA Saclay \cite{lecturesSaclay}, and in a talk at the Simons Foundation \cite{talkSimons}. SR thanks Riccardo Guida for the kind invitation to deliver the Saclay lectures.

PK is supported by DOE grant DE-SC0009988 and by the Corning Glass Works Foundation Fellowship Fund at the Institute for Advanced Study. The work of SR and JQ is supported by the Simons Foundation grant 488655 (Simons Collaboration on the Nonperturbative Bootstrap). SR is supported by Mitsubishi Heavy Industries as an ENS-MHI Chair holder.

\appendix
\section{Lorentzian 4pt correlator with no convergent OPE channel}
\label{app:lorentz}
In this section we will give an example of a Lorentzian 4pt configuration in which there's no convergent OPE channel. For simplicity let's consider the correlators of identical scalar operators. Recall that, in a general QFT, Lorentzian correlators can be recovered from Euclidean correlators by analytic continuation. Starting from a configuration of Euclidean points $x_i=(\tau_i,\mathbf{x}_i)$ with ordered times 
\be
\label{eq:or0}
\tau_1>&\tau_2>\ldots>\tau_n\,,
\ee
we analytically continue each time variable as $\tau_i = \eps_i+i t_i$ and take the limit $\eps_i\to0$, preserving the ordering of real parts. The result is interpreted as the Lorentzian correlator at (Lorentzian) points $y_i=(t_i,\mathbf{x}_i)$. Schematically:
\begin{equation}
\begin{split}
\<0|\phi(t_1,\mathbf{x}_1)\ldots\phi(t_n,\mathbf{x}_n)|0\>:=\lim\limits_{\substack{\eps_i\rightarrow0 \\\eps_1>\ldots>\eps_n}}\<\phi(\eps_1+it_1,\mathbf{x}_1)\ldots\phi(\eps_n+it_n,\mathbf{x}_n)\>
\end{split}
\end{equation}
Now we will apply this to a 4pt function in a CFT. In a CFT, this analytic continuation can be performed starting from Eq.~(\ref{eq:4pt}). We just complexify all Euclidean times as described above, and then take the limit. It is easy to see (exercise) that the distances $x_{ij}^2$ do not vanish in this process, except perhaps at the very end if the Lorentzian points $y_i$ are lightlike separated. We will be interested in the case when all points are spacelike or timelike separated. So the prefactor in Eq.~(\ref{eq:4pt}) is thus analytically continued (notice that there is an interesting phase for timelike separation).

In order to analytically continue the factor $g(u,v)$, we will use the existence of the conformal block expansion (\ref{eq:cbexpansion}) which as mentioned there is convergent for $|\rho|,|\bar{\rho}|<1$ (``OPE convergence region''). Concretely, we are instructed to compute $u,v$ corresponding to complexified Euclidean times, then evaluate $z,\bar{z}$ defined by~\eqref{eq:zzbar0}, which gives 
\be
z,\bar{z}=\frac{1}{2}(1+u-v\pm\sqrt{(1+u-v)^2-4u}),
\ee
then evaluate the corresponding $\rho,\bar\rho$ via~\eqref{eq:rho}, and finally stick these into the expansion (\ref{eq:cbexpansion}). This procedure defines an analytic function of $\tau_i$ as long as $|\rho|,|\bar{\rho}|<1$.\footnote{Note that even though $z,\bar{z}$ will have a branch point when $(1+u-v)^2-4u=0$, the function $g(u,v)$ is symmetric under the intercharge of $z,\bar{z}$ and will remain analytic as a function of complexified Euclidean times.} The question then is if this condition will hold all along the analytic continuation curve needed to recover the Lorentzian correlator, including the endpoint. If this happens, Lorentzian correlator can be computed by summing up a convergent expansion, in particular it is non-singular. 

Above we describe how to use the $s$-channel expansion for the analytic continuation. A priori we can also use the $t$- and $u$-channels for this purpose, starting from the $t$- and $u$-channel versions of Eq.~(\ref{eq:4pt}): 
\begin{equation}
\label{eq:4ptt}
\<\f(x_1)\f(x_2)\f(x_3)\f(x_4)\>=\frac{1}{(x_{23}^2)^{\De_\f}(x_{14}^2)^{\De_\f}}g(u_t,v_t)  =\frac{1}{(x_{13}^2)^{\De_\f}(x_{24}^2)^{\De_\f}}g(u_u,v_u).
\end{equation}
The cross ratios $u_t,v_t$ are obtained from $u,v$ via $x_1\leftrightarrow x_3$, and $u_u,v_u$ via $x_2\leftrightarrow x_3$. The functions $g(u_t,v_t)$, $g(u_u,v_u)$ can be computed via the corresponding conformal block expansions with their own regions of analyticity set by the conditions $|\rho_t|,|\bar{\rho}_t|<1$ and $|\rho_u|,|\bar{\rho}_u|<1$. 

It is not a priori clear and requires a separate analysis, which OPE channel, if any, is convergent for a given Lorentzian configuration. The answer turns out to depend, generically, only on the causal structure of the configuration (who is timelike, who is spacelike). 
The OPE can stop converging in two ways: either at the end point of the analytic continuation, or somewhere along the way. As we will show in~\cite{paper2}, for the $s$-channel we always have $|\rho|,|\bar\rho|\le 1$, so OPE converges along the way but may diverge at the end point. For other channels the OPE may start diverging already along the way.

We will give an exhaustive discussion of these phenomena, for all possible causal structures, in a later publication~\cite{Jiaxin}. Here we will just give an extreme example of a configuration where all channels diverge. 

Consider the causal ordering 
\be
\label{eq:ord}
y_3\rightarrow y_1\rightarrow y_4\rightarrow y_2\,,
\ee
where $y_i\to y_j$ means that $y_i$ is in the past open lightcone of $y_j$. We pick some points $(i t_i, \mathbf{x}_i)$ corresponding to this causal ordering, as well as some initial Euclidean times $\eps_i$ satisfying the ordering $\eps_1>\eps_2>\eps_3>\eps_4$, and consider a curve of complexified points corresponding to these initial and final positions. E.g.~we can use linear interpolation:
\be
x_i(\theta) = ( (1-\theta)\eps_i + \theta it_i , \mathbf{x}_i),\qquad \theta\in[0,1]\,.
\ee
We choose the initial point with $|\rho|,|\bar\rho|<1$, and we would like to see if this condition stays true along this curve. For this it is enough to evaluate $z,\bar z$ and see if they cross the cut $[1,+\infty)$ which corresponds to $|\rho|=1$. This is how the check is carried out in practice for the $s$-channel. For the $t$- and $u$-channel, we have the same check in terms of $z_t,\bar{z}_t$ and $z_u,\bar{z}_u$. But in fact we have relations 
\be
z_t= 1-z,\qquad z_u= 1/z
\ee
and similarly for $\bar z$. These relations map the $[1,+\infty)$ cut on $(-\infty,0]$ and $[0,1]$, respectively.
Thus we don't have to redo the analysis for $z_t,\bar{z}_t$ and $z_u,\bar{z}_u$ separately, we just have to watch if the $s$-channel $z,\bar z$ crosses these additional cuts to conclude about the convergence of the $t$- and $u$-channel OPEs.

\begin{figure}[H]
	\centering
	\includegraphics[scale=0.6]{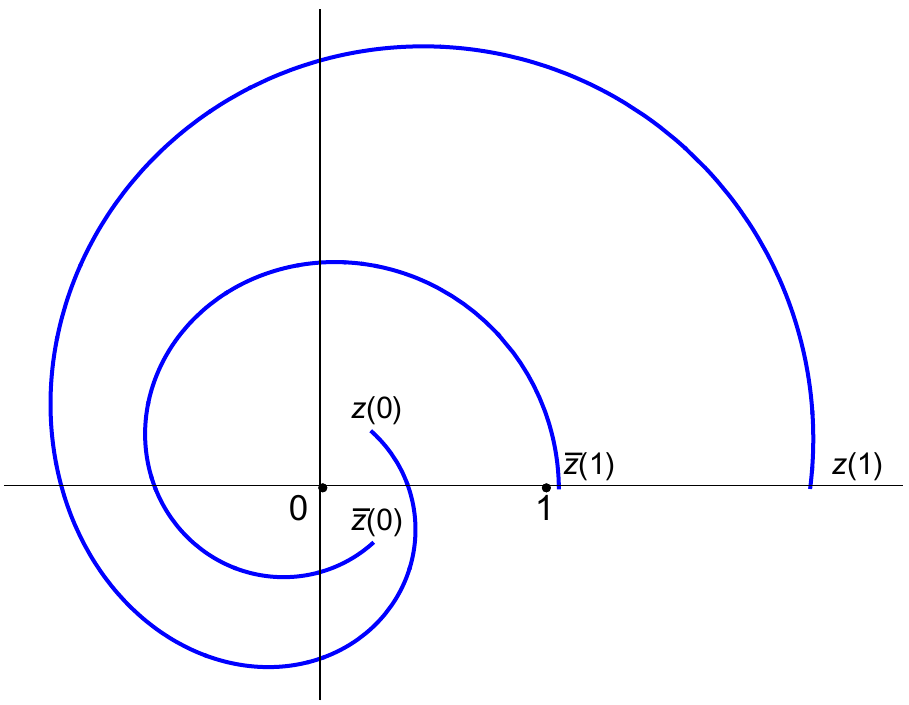}
	\caption{\label{plot-orbit3142}The curves $z(\theta)$ and $\bar{z}(\theta)$ for the causal ordering $y_3\rightarrow y_1\rightarrow y_4\rightarrow y_2$.}
\end{figure}

In practice, we just pick some numerical values for the initial and final points (respecting the orderings), plot the curves $z(\theta)$, $\bar{z}(\theta)$ and see what they do. For the causal ordering~\eqref{eq:ord}, we get the plot shown in Fig.~\ref{plot-orbit3142}. To draw the plot we picked numerical values:
\be
\begin{array}{llll}
\eps_1=4\,, &\eps_2=3\,, &\eps_3=2\,, &\eps_4=0\,,\\ 
y_1 =(2,0,0,0)\,,\ &y_2 =(20,0,0,0)\,,\ &y_3 =(0,0.9,0,0)\,,\ &y_4 =(3,0,0,0)\,,
\end{array}
\ee
where $y_i=(t_i,{\bf x}_i)$. Any other initial point $\eps_1>\eps_2>\eps_3>\eps_4$ and the final point corresponding to the ordering~\eqref{eq:ord} gives rise to a topologically equivalent configuration of curves.
We see that the curves $z(\theta)$, $\bar z(\theta)$ touch the $[1,\infty)$ cut at $\theta=1$ but do not cross it at the intermediate values of $\theta$. This means that $|\rho|=|\bar{\rho}|=1$ at the corresponding Lorentzian configuration. Furthermore, both curves cross the $t$-channel cut $(-\infty,0]$, which according to the above discussion means $|\rho_t|>1$, $|\bar\rho_t|>1$. One of the two curves also crosses the $u$-channel cut $[0,1]$, which means $|\rho_u|>1$. We conclude that the Lorentzian configuration under study is outside the region of OPE convergence of any of the three channels.

The given recipe to determine which channels diverge would require some care in situations when a curve crosses a cut and then goes back, or when the $z(\theta)$ and $\bar{z}(\theta)$ curves cross the same cut in opposite directions. We will discuss these subtleties and their interpretation in~\cite{Jiaxin}. In the given example they do not occur, so our conclusion that all three channels diverge is robust.

Another comment is in order concerning the 2d CFT case. In this case, the region of analyticity of 4pt functions is larger than $|\rho|, |\bar{\rho}|<1$, being instead given by the condition $|q|, |\bar{q}|<1$~\cite{Hogervorst:2013sma} where $q$ is Al.~Zamolodchikov's uniformizing variable.  Using this variable, one can show that the Lorentzian 4pt function in a 2d CFT is analytic for all possible causal orderings away from null cone singularities~\cite{Maldacena:2015iua}.

\section{Proof of lemma~\ref{lemma:functions}}
\label{app:lemma}
	
To prove the first inequality,\footnote{See~\cite{Schwarz}, Exercise 6.3 for similar arguments. For this result it's only important that $|\phi(w)|\le 1$. That it's one-to-one and avoids the cut does not matter.} we start by constructing a map $\tl\varphi(w)$ from $\mathbb{D}$ into $\mathbb{D}$ which satisfies $\tl\varphi(0)=0$. This is achieved by a fractional linear transformation as follows
\be
\label{eq:FLT}
\tl\varphi(w)=\frac{\varphi(w)-\varphi(0)}{1-\varphi(w)\overline{\varphi(0)}}.
\ee
Now, Schwarz lemma implies that $|\tl\varphi(w)|\leq |w|$ and so $1-|\tl\varphi(w)|\geq 1-|w|$. At the same time, we find 
\be
1-|\tl\varphi(w)|^2=\frac{(1-|\varphi(w)|^2)(1-|\varphi(0)|^2)}{(1-\varphi(w)\overline{\varphi(0)})(1-\varphi(0)\overline{\varphi(w)})}\leq C(1-|\varphi(w)|)\,,\qquad C=2\frac{1+|\varphi(0)|}{1-|\varphi(0)|}
\ee
where the first equality follows by a short computation from~\eqref{eq:FLT}, and to get the inequality we bounded some factors using $|\varphi(w)|\le 1$. Furthermore, since $1-|\tl\varphi(w)|^2=(1-|\tl\varphi(w)|)(1+|\tl\varphi(w)|)\geq 1-|\tl\varphi(w)|$, we find
\be
1-|w|\leq 1-|\tl\varphi(w)|\leq1-|\tl\varphi(w)|^2\leq  C(1-|\varphi(w)|).
\ee

To prove the second inequality, it will be important that $\varphi(w)$ is one-to-one and that $\varphi(w)\ne 0$.\footnote{It won't be important that it avoids the rest of the cut.} Under these conditions the function $\frac{1}{\varphi(w)}$ is holomorphic and one-to-one. Such functions from $\mathbb{D}$ onto a subset of $\C$ are called univalent, or schlicht~\cite{Duren}. The shifted and rescaled function
\be
h(w)=-\frac{\varphi(0)^2}{\varphi'(0)}\p{\frac{1}{\varphi(w)}-\frac{1}{\varphi(0)}},
\ee
is then also univalent, and in addition satisfies normalization conditions $h(0)=0$ and $h'(0)=1$. A basic result about normalized univalent functions is the Growth Theorem (\cite{Duren}, Theorem 2.6)
\be
|h(w)|\leq \frac{|w|}{(1-|w|)^2}.
\ee
This immediately implies the second bound in~\eqref{eq:Lind-bound}.

\section{Comments on the proof of theorem~\ref{thm:vlad2}}
\label{app:proofVlad2}

Compared to theorem~\ref{thm:vlad}, theorem~\ref{thm:vlad2} has only two essentially new ingredients. First, we now have the freedom of choosing $v\in V$ so we want to show that this choice doesn't matter, and second, we have to prove that the boundary value is holomorphic in $w$. Without these two ingredients, the proof of section~\ref{sec:proof} goes through without any essential modifications. 

Let us briefly recall the main steps of that proof, but now in the context of theorem~\ref{thm:vlad2}.\footnote{Our proof is an adaptation of the proof of theorem 7.2.6 in~\cite{RealSubmanifolds}.} First, for a Schwartz test function $f(x)$ we define
\be
	L_v(w,\e) = \int d^dx g(w,x+iv\e) f(x).
\ee
Using integration by parts, we show that
\be\label{eq:integrationbypartsidentity}
	\ptl_\e^k L_v(w,\e) = (-i)^k\int d^dx g(w,x+iv\e)v^{\mu_1}\cdots v^{\mu_k} \ptl_{\mu_1}\cdots\ptl_{\mu_k}f(x).
\ee
We then use this identity and the slow-growth condition on $g$~\eqref{eq:slow-growth2} to bound
\be
	|\ptl_{\e}^k L_v(w,\e)| \leq \frac{C_k}{\e^{2K}}.
\ee
for some $C_k>0$ that is proportional to some semi-norm of $f$. In what follows, it will be important to us how $C_k$ depends on $v$. It is easy to see that
\be
	|\ptl_{\e}^k L_v(w,\e)| \leq \frac{C_k'||v||_\infty^k ||v||_2^{-2K}}{\e^{2K}}.
\ee
for some $C_k'>0$ that is independent of $v$. Furthermore, since the bound~\eqref{eq:slow-growth2} is independent of $w$, $C'_k$ is also independent of $w$.\footnote{This holds on compact subsets $\cK\subset U$, see footnote~\ref{ft:compactfootnote}.} Then we use the obvious analogue of~\eqref{eq:NLformula} starting from sufficiently large $k$ to conclude 
\be\label{eq:improvedbound}
	|\ptl_{\e} L_v(w,\e)| \leq {C||v||_\infty^k ||v||_2^{-2K}}
\ee
for some $C>0$ proportional to a semi-norm of $f$. This immediately implies that 
\be
	L_v(w,\e)=-\int_\e^{\e_0}\ptl_{\e} L_v(w,\e)+ L_v(w,\e_0)
\ee
is continuous down to $\e=0$ and that thus defined $L_v(w,0)$ depends continuously on $f$ in $\cS(\R^d)$. The slight refinements that we made to the bound~\eqref{eq:improvedbound}, i.e.\ observing that it holds uniformly in $w$ and exhibiting its dependence on $v$, allow us to make the following statement: the limit $L_v(w,\e)\to L_v(w,0)$ is reached uniformly on compact sets $\cK\subset U$ in $w$ and on compact sets $\cV\subset V$ in $v$ (recall that $V$ doesn't contain $0$). This statement is the key in proving that the limit is independent of $v\in V$ and is holomorphic in $w$.

The fact that $L_v(w,0)$ is holomorphic in $w$ is now indeed straightforward, since $L_v(w,\e)$ is holomorphic in $w$ for $\e>0$.\footnote{The standard argument is as follows. Suppose holomorphic functions $h_n$ converge uniformly to some function $h$. Then, first of all, $h$ is continuous because $h_n$ are and the limit is uniform. Second, the uniform limit can be exchanged with contour integration. Since integrals of $h_n$ over closed curves are $0$, so are the integrals of $h$. By Morera's theorem, this implies holomorphicity of $h$.} To prove that it is independent of $v$ requires a bit more work. Take $v_1,v_2\in V$ and write
\be
	L_{v_1}(w,\e)-L_{v_2}(w, \e)&=\int d^dx (g(w,x+iv_1\e)-g(w,x+iv_2\e)) f(x)\nn\\
	& = \int d^dx \int_0^1 dt \,\ptl_t g(w,x+iv(t)\e) f(x), \qquad v(t)=t v_1+(1-t)v_2\nn\\
	& = -i\e \int_0^1 dt \int d^dx g(w,x+iv(t)\e) \,\,(v_1-v_2)\cdot\ptl f(x)\nn\\
	& = -i\e \int_0^1 dt \,\tl L_{v(t)}(w,\e).
\ee
where $\tl L_v(w,\e)$ is defined as $L_v(w,\e)$ but with $(v_1-v_2)\cdot\ptl f(x)$ instead of $f(x)$. Since $(v_1-v_2)\cdot\ptl f(x)$ is also a test function, we have that by the same arguments as the above,
$\tl L_v(w,\e)$ converges to a finite limit $\tl L_v(w,0)$ uniformly in $v$ on compacts of $V$. This implies that the integral
\be
	\int_0^1 dt \,\tl L_{v(t)}(w,\e)
\ee
has a finite limit as $\e\to 0$, and thus
\be
	L_{v_1}(w,\e)-L_{v_2}(w, \e)= -i\e \int_0^1 dt \,\tl L_{v(t)}(w,\e)\to 0.
\ee

\small

\bibliography{lorentz}
\bibliographystyle{utphys}

\end{document}